\documentclass[twocolumn]{openjournal}

\graphicspath{{./}{plots/}}
\accepted{\today}
\shortauthors{Richardson \& Plazas Malag\'on et al.}

\usepackage{graphicx}
\usepackage{hyperref}
\usepackage{amsmath}
\usepackage{color}
\usepackage[caption=false]{subfig}

\begin{document}

\title{The CosmoQuest Moon Mappers Community Science Project: The Effect of Incidence Angle on the Lunar Surface Crater Distribution}
\email{aplazas@astro.princeton.edu}

\author{Matthew Richardson}
\affil{Planetary Science Institute, 1700 East Fort Lowell, Suite 106, Tucson, AZ 85719, USA}
\author{Andr\'es A. Plazas Malag\'on}
\affil{Department of Astrophysical Sciences, Peyton Hall, Princeton University, Princeton, NJ 08544, USA}
\affil{Astronomical Society of the Pacific, 390 Ashton Avenue, San Francisco, CA 94112, USA}

\author{Larry A. Lebofsky}
\affil{Planetary Science Institute, 1700 East Fort Lowell, Suite 106, Tucson, AZ 85719, USA}

\author{Jennifer Grier}
\affil{Planetary Science Institute, 1700 East Fort Lowell, Suite 106, Tucson, AZ 85719, USA}

\author{Pamela Gay}
\affil{Planetary Science Institute, 1700 East Fort Lowell, Suite 106, Tucson, AZ 85719, USA}
\affil{Astronomical Society of the Pacific, 390 Ashton Avenue, San Francisco, CA 94112, USA}

\author{Stuart J. Robbins}
\affil{Southwest Research Institute, 1050 Walnut St., Suite 300, Boulder, CO 80302, USA}

\author{The CosmoQuest Team}

\newcommand{\mm}{{\tt{Moon Mappers}}}

\newcommand{\subf}[2]{%
  {\small\begin{tabular}[t]{@{}c@{}}
  #1\\#2
  \end{tabular}}%
}

\begin{abstract}
 The CosmoQuest virtual community science platform facilitates the creation and implementation of astronomical research projects performed by citizen scientists. One such project---called \mm---aides in determining the feasibility of producing crowd-sourced cratering statistics of the Moon's surface. Lunar crater population statistics are an important metric used to understand the formation and evolutionary history of lunar surface features, to estimate relative and absolute model ages of regions on the Moon's surface, and to establish chronologies for other planetary surfaces via extrapolation from the lunar record. It has been suggested and shown that solar incidence angle has an effect on the identification of craters, particularly at meter scales. We have used high resolution image data taken by the Lunar Reconnaissance Orbiter’s Narrow Angle Camera of the \emph{Apollo} 15 landing site over a range of solar incidence angles and have compiled catalogs of crater identifications obtained by minimally-trained members of the general public participating in CosmoQuest's \mm\ project. We have studied the effects of solar incidence angle spanning from $\sim 27.5^{\circ}$ to $\sim 83^{\circ}$ (extending the incidence angle range examined in previous works), down to a minimum crater size of 10 m, and find that the solar incidence angle has a significant effect on the crater identification process, as has been determined by subject-matter experts in other studies. The results of this analysis not only highlight the ability to use crowd-sourced data in reproducing and validating scientific analyses, but also indicate the potential to perform original research.

\end{abstract}



\section{Introduction}

A fundamental tool of planetary geology is the statistics of impact craters.  The size-frequency distribution of craters is the only\footnote{{Other than \emph{in situ} methods, \emph{e.g.,} \citet{farley14}.}} method available on solid bodies other than Earth to model surface ages \citep{bell20,neukum01}. On the Moon, calibrations provided by \emph{Apollo} and \emph{Luna} samples allow us to calculate the lunar crater calibration function, which relates absolute age of the sample with the spatial density of craters larger than 1 km \citep{neukum94}. Since many surfaces require measuring craters either smaller than 1 km due to their small area or youth, or measuring only larger craters due to their older age and crater saturation at smaller sizes, a model crater size-frequency distribution (SFD) must be used to estimate the 1 km and larger crater population.  However, ensuring that the craters measured are a representative sample of the craters on that surface is not as straightforward as it might seem. 

Identification of impact craters, with statistics typically given in the form of an SFD, is not trivial. Factors such as the large number of craters of different diameters on the lunar surface and the individual variations in crater identification from different researchers make this a challenging task. 

\begin{figure}[htp]
    \centering
    \subfloat{
        \includegraphics[width=0.9\columnwidth]{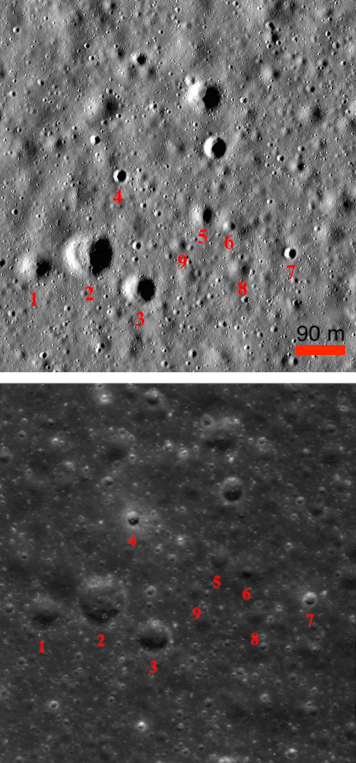}
    }

    \caption{Portion of LROC-NAC images {\tt{M146959973LE}} (upper panel) and {\tt{M109215691LE}} over the same lunar region and at different solar incidence angles with respect to the normal (77.5$^{\circ}$ and 27.5$^{\circ}$, respectively). Corresponding craters are marked with numbers by one of the authors (L.L). {{The scale is such that the diameter of crater number 2 corresponds to approximately 90 m.}}Crater detection and identification is affected by the different illumination angles.}  
    \label{fig:angles}
\end{figure}

Another important parameter that plays a central role when measuring the SFD of craters on the lunar surface is the solar incidence angle \citep{young75,wilcox05}, defined as the angle between the normal to the surface and the position vector of the Sun on the sky. Images that are taken under different illumination conditions highlight different characteristics on the lunar surface (Fig.~\ref{fig:angles}). Topographic features of craters such as rims generate more pronounced shadows at higher incidence angles (\emph{i.e.}, with the solar illumination nearer the local horizon), facilitating the identification of craters in some cases \citep{head72,soderblom72,young75}, although potentially hiding small craters under the shadows of larger craters in other cases \citep{moore72}. On the other hand, rays formed from ejecta of primary and secondary craters {{could}} have a distinct brightness in optical light, so higher sun that emphasizes these brightness differences {{may}} aid in identification of those craters \citep{neish13,elliot18}, while making difficult the identification of other craters \citep{wilcox05}.  

The precise determination of the crater equilibrium state---in which craters smaller than the equilibrium diameter are being produced at the same rate at which they are being destroyed---is fundamental to the accuracy of crater-based age determinations and regolith depth estimations \citep{shoemaker70,ostrach11,xiao15}. The equilibrium diameter can be identified as a break in the slope in a cumulative SFD plot \citep{gault70,schultz77}. \citet{hirabashi17} demosntrate, for example, that the  the \emph{Apollo} 15 landing site is in equilibrium at crater sizes below approximately 100 m.

{{Previous work (\emph{e.g.}, \cite{soderblom72,wilcox05})}} suggests that the determination of the equilibrium state is affected considerably by the different crater number counts found as a consequence of different solar incidence angles, a hypothesis that is also consistent with the analysis of lunar data by \citet{ostrach11}.  {{\citet{schultz77} (Fig. 3) have also demonstrated the relative difference between crater counts as a function of solar incidence angles}}.

In this paper, we use the high resolution data of the \emph{Apollo} 15 landing region imaged by the Lunar Reconnaissance Orbiter \citep{robinson10} Narrow-Angle Camera (LROC NAC) \citep{chin06} to study the effect of solar incidence angle on crater counts in overlapping regions from different images. The LROC camera system has imaged the lunar surface several times at sub-meter resolution over a wide range of illumination angles. 

Lunar craters are marked by minimally-trained non-subject matter volunteers via the {\tt{Moon Mappers}}\footnote{\url{https://cosmoquest.org/x/science/moon/}} community science portal, a part of the CosmoQuest\footnote{\url{https://cosmoquest.org/x/}} project \citep{gay14,robbins12}. 
\citet{robbins14} have demonstrated that community science volunteers produce, on average, results comparable to those by subject-matter experts when identifying lunar craters. Earlier works by the CosmoQuest team analyzing similar data annotated by the public have corroborated the dependence of crater counts on solar incidence angle \citep{antonenko13,grier18}. 

The paper is organized as follows. In \S 2 we describe the LROC data used for this analysis. Section \S 3 reviews the CosmoQuest project and the \mm\ interface, detailing the crater-marking process by the volunteers, and the analysis performed on these data to determine the craters SFDs. We report our results in \S 4, and conclude in \S 5.

\section{Data}
\label{sec:data}

\begin{figure*}[htp]
    \centering
    \subfloat{
        \includegraphics[height=8cm]{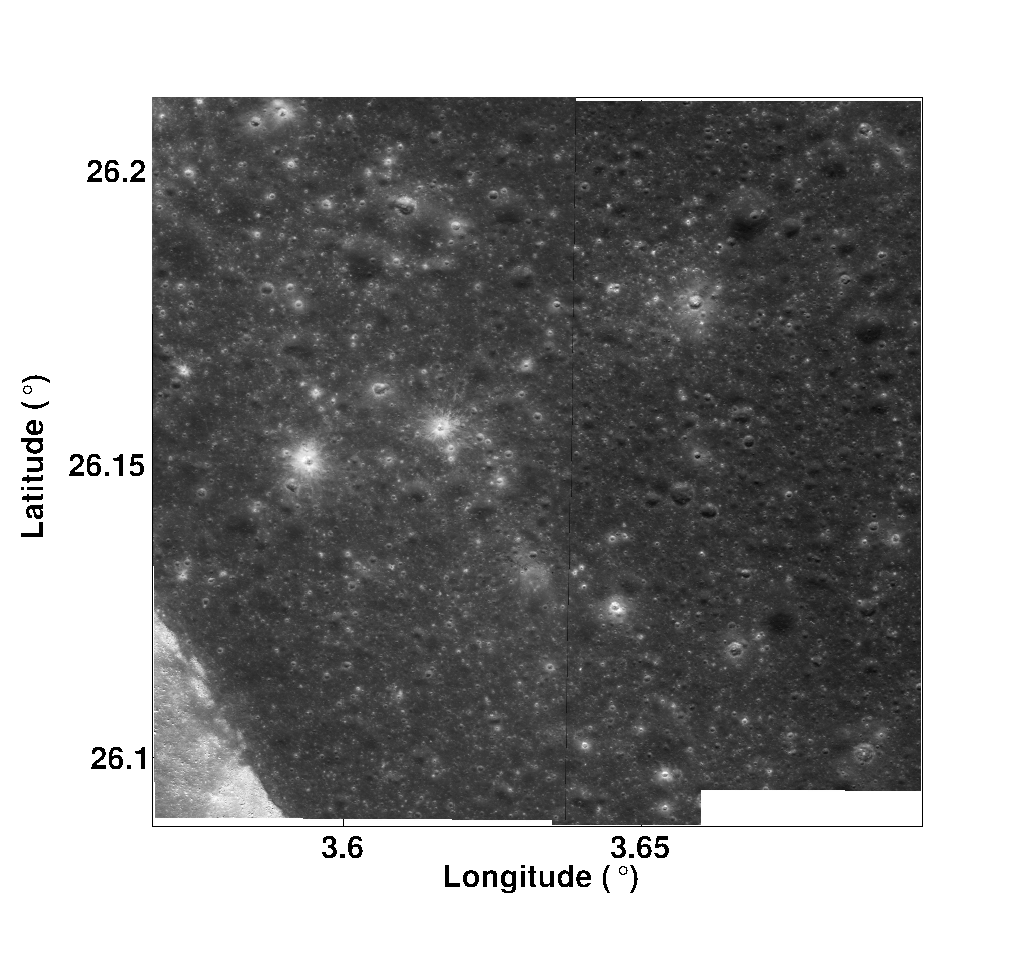}
    }
    \subfloat{
        \includegraphics[height=8cm]{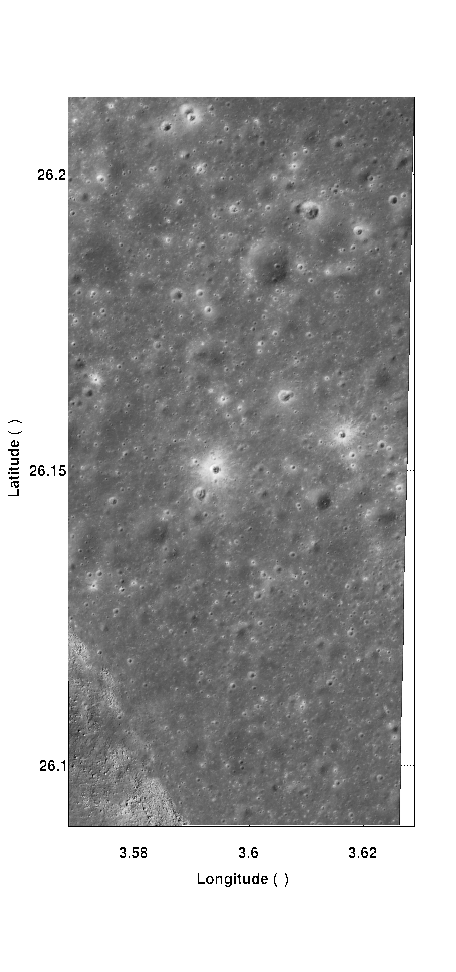}
    }
    \subfloat{
        \includegraphics[height=8cm]{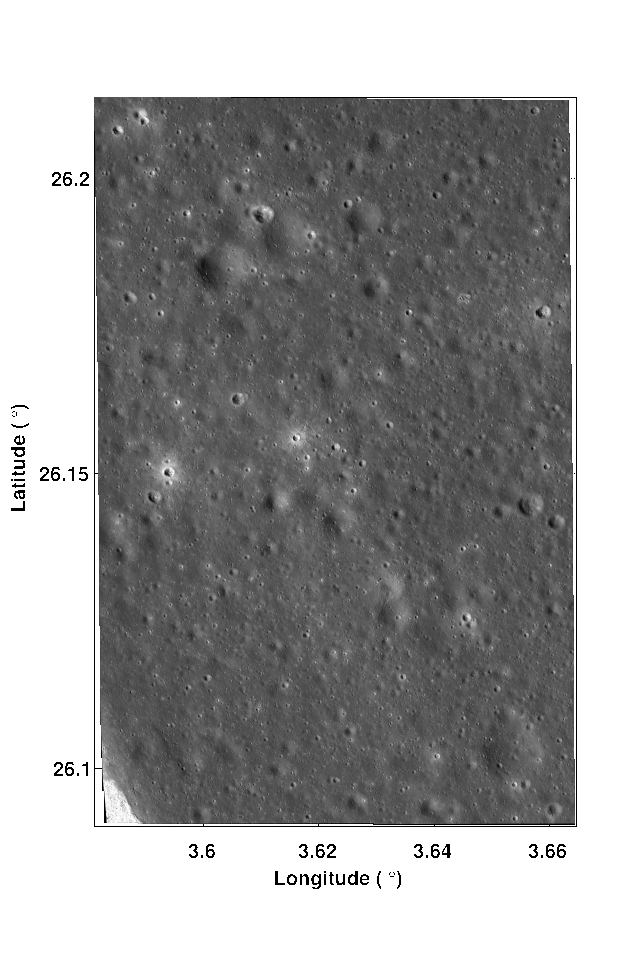}
    }
    
    \subfloat{
        \includegraphics[height=8cm]{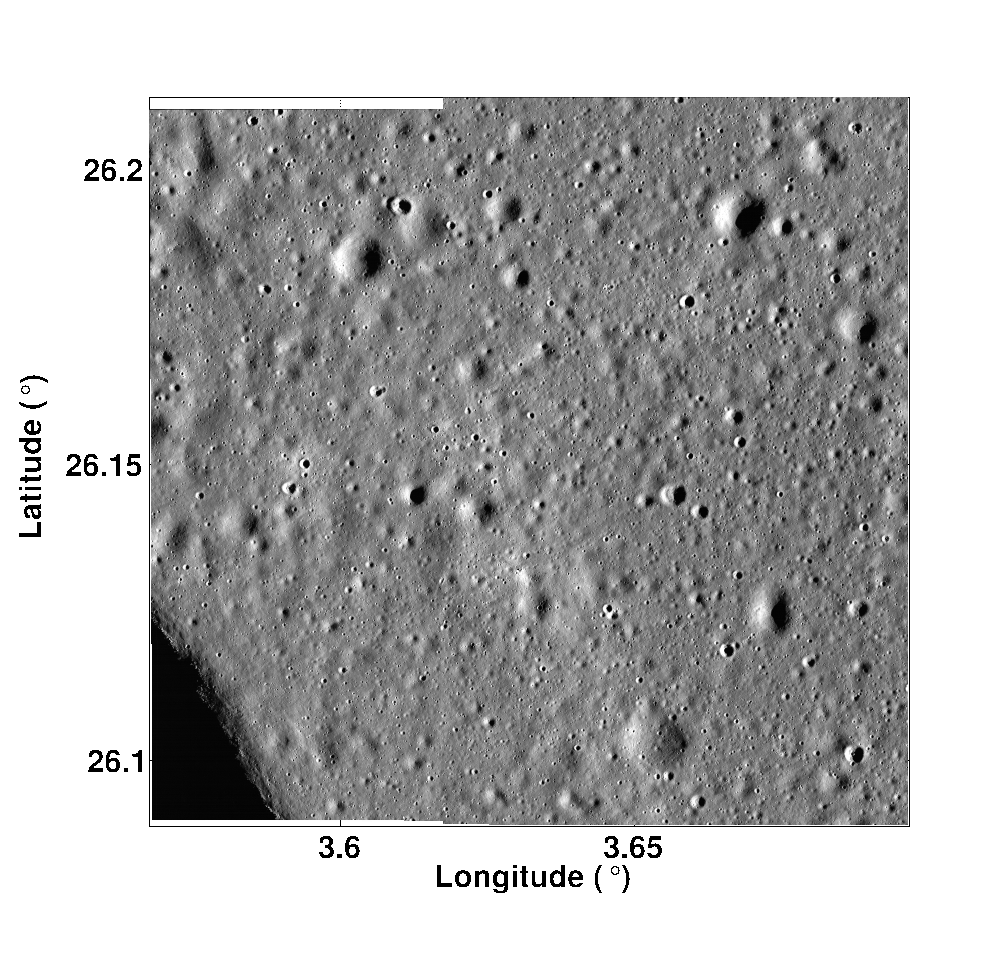}
    }
    \subfloat{
        \includegraphics[height=8cm]{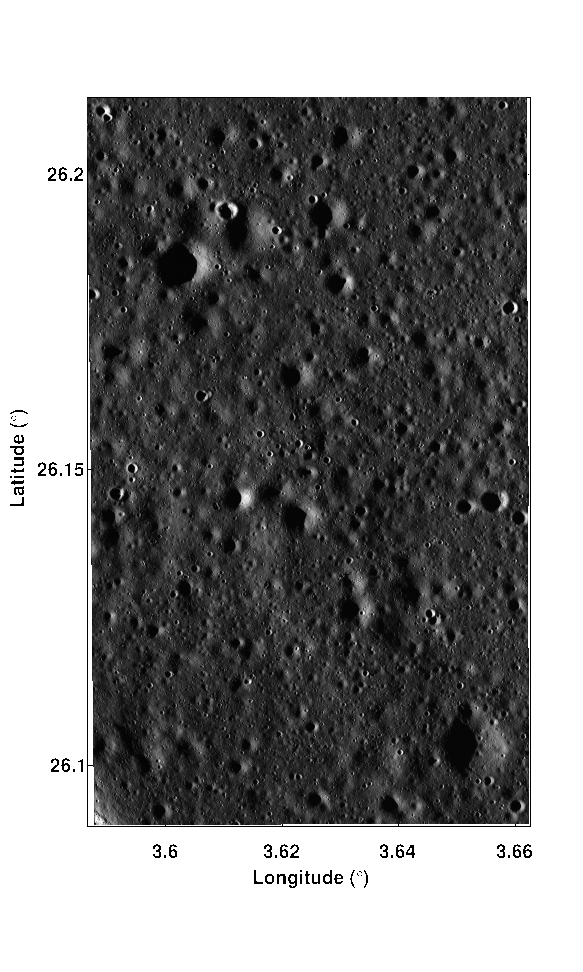}
    }

    \caption{LRO NAC images at each incidence angle used for the analysis in this work. The axes are in selenographic coordinates. From top to bottom and left to right: {\tt {M109215691LERE}} (27.5$^{\circ}$), {\tt {M111578606RE}} (38.0$^{\circ}$), {\tt {M119829425LE}} (58.0$^{\circ}$), {\tt {M146959973LERE}} (77.5$^{\circ}$), and {\tt {M117467833RE}} (83.0$^{\circ}$). Data associated with image names ending with ``LE", ``RE", and ``LERE" correspond to crater data derived from the left, right, and both NAC cameras, respectively. {Crater number 2 of Figure \ref{fig:angles} with an approximate diameter of 90 m is located at coordinates (longitude, latitude) $\approx (3.66^{\circ}, 26.14^{\circ})$}}. 
    \label{fig:images_angles}
\end{figure*}

The data used for this analysis consist of seven overlapping images acquired by the LROC NAC. The overlap region used between the images has an estimated area of 14.76 km$^{2}$ and encompasses the \emph{Apollo} 15 landing site. The images have an average scale of $\approx$ 0.5 m per pixel, and they were imaged at solar incidence angles of 27.5$^{\circ}$, 38.0$^{\circ}$, 58.0$^{\circ}$, 77.5$^{\circ}$, and 83$^{\circ}$ (Fig.~\ref{fig:images_angles}; some NAC left-right pairs are included, which is why there are only five incidence angles despite our use of seven images). This range of incidence angles extends the angle ranges examined in previous works (e.g., \citet{wilcox05,ostrach11}). Many studies---some similar in nature to our analysis---have been conducted within this region of the lunar surface; it serves as an ideal location to test the effects of incidence angle on crowd-sourced crater identifications and measurements. Other community science projects, such as {\tt{Moon Zoo}}, performed similar analyses to the one presented in this work using images from different \emph{Apollo} mission landing sites \citep{joy11,bug16}.

\begin{figure*}[htp]
\centering
\resizebox{0.7\hsize}{!}{\includegraphics{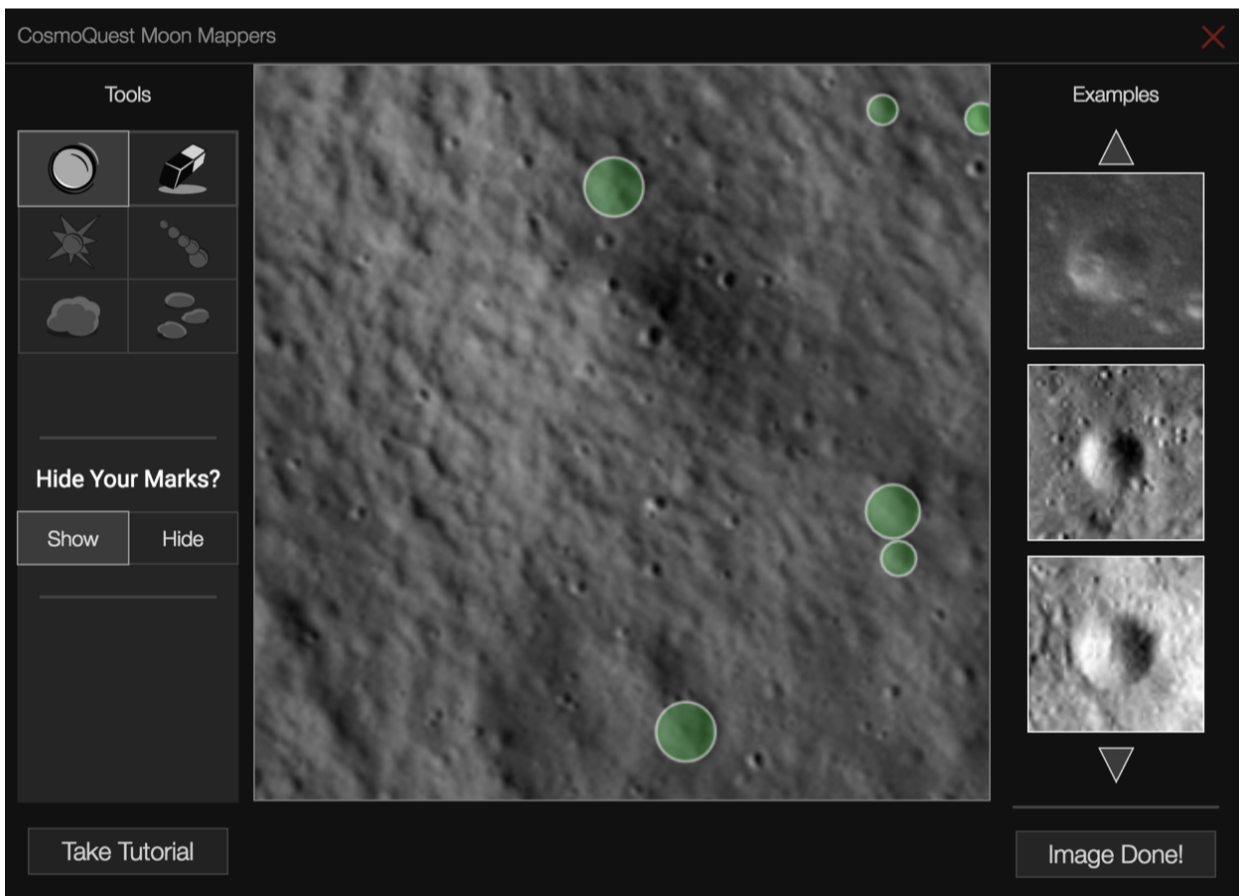}}

\caption{\mm\ interface used by members of the general public to annotate craters. After completing a brief tutorial on how to use the tool, the interface guides volunteers to mark craters by drawing circles around the crater rim. The tool guides them to only select craters of a certain minimum size. The interface also has a bar on the right that contains examples of craters in all stages of degradation to assist volunteers in crater identification. {{The images presented to the volunteers have a linear stretch with a range of $\pm$ 0.5 \% (users did not have the ability to change the stretch).}}}
\label{fig:interface}
\end{figure*}

All image data were downloaded from the Planetary Data Systems (PDS)\footnote{\url{https://pds.nasa.gov}} LROC image node. The images were processed using the Integrated Software for Images and Spectrometers (ISIS)\footnote{\url{https://isis.astrogeology.usgs.gov/}} software package and projected to a local Mercator projection. Our region of interest focuses on most of the area between 26.09$^{\circ}$ to 26.21$^{\circ}$ latitude and 3.56$^{\circ}$ to 3.70$^{\circ}$ longitude, which roughly translates to a grid of 3.64$\times$4.24 km$^{2}$. As a single image easily blankets most of the target region, even at full screen resolution, the imaged area is far too large to identify craters at the ten-meter scale. Furthermore, attempting to identify craters present within an image would have the appearance of being a daunting task. To make the identification of craters manageable and enhance the feasibility of identifying craters at the minimum desired crater diameter, images were sliced into sub-images 450 pixels across and tall, with overlaps. To ensure accuracy in the georeferencing of positions within the images, all images were uploaded into the Java Mission-planning and Analysis for Remote Sensing (JMARS)\citep{christensen09} software package to estimate the selenographic corner coordinates of the images. The tables in the appendix show the crater population results used in this work. 

\begin{figure*}[htp]
    \centering
    \subfloat{
        \includegraphics[width=0.99\columnwidth]{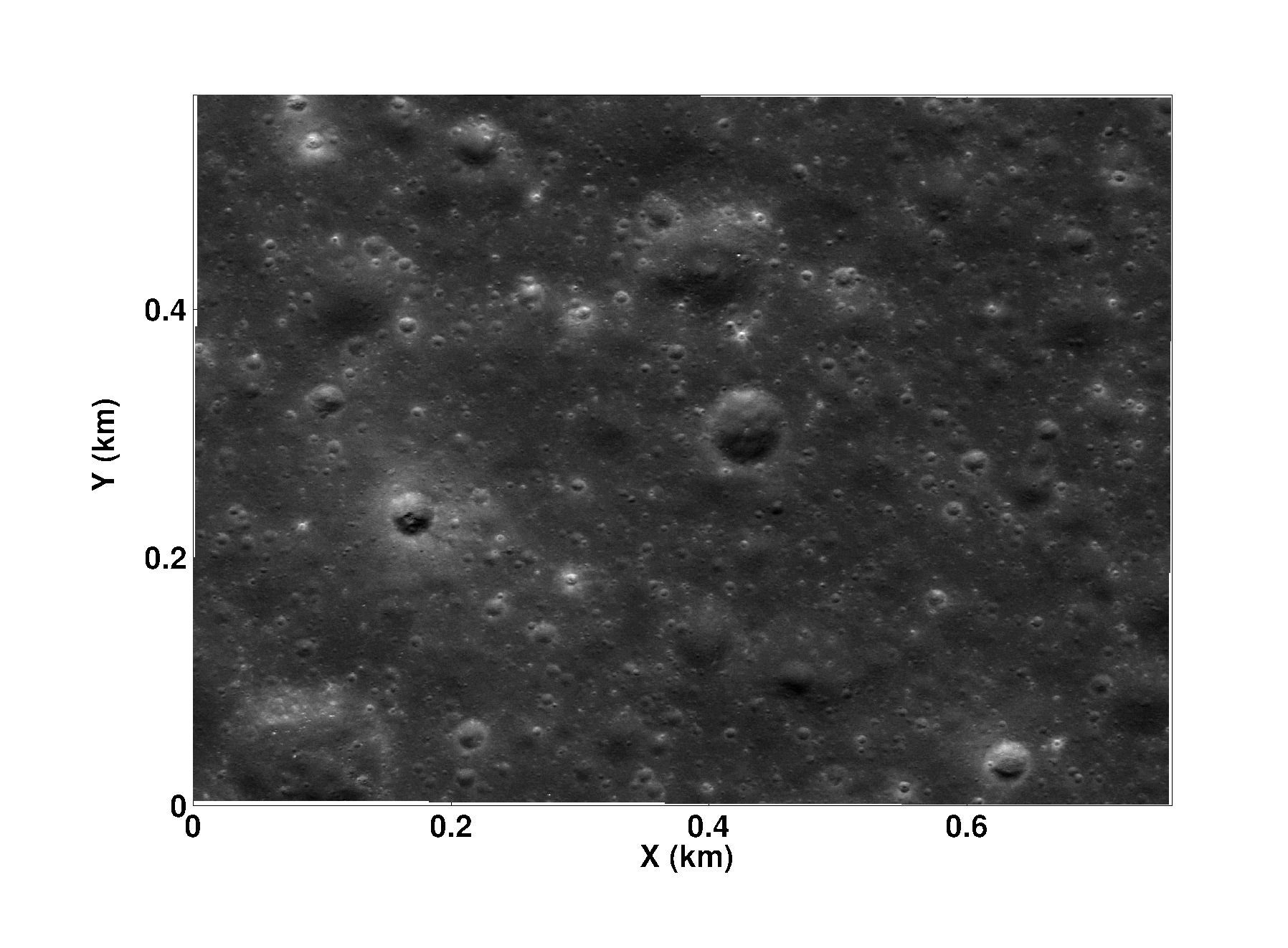}
    }
    \subfloat{
        \includegraphics[width=0.99\columnwidth]{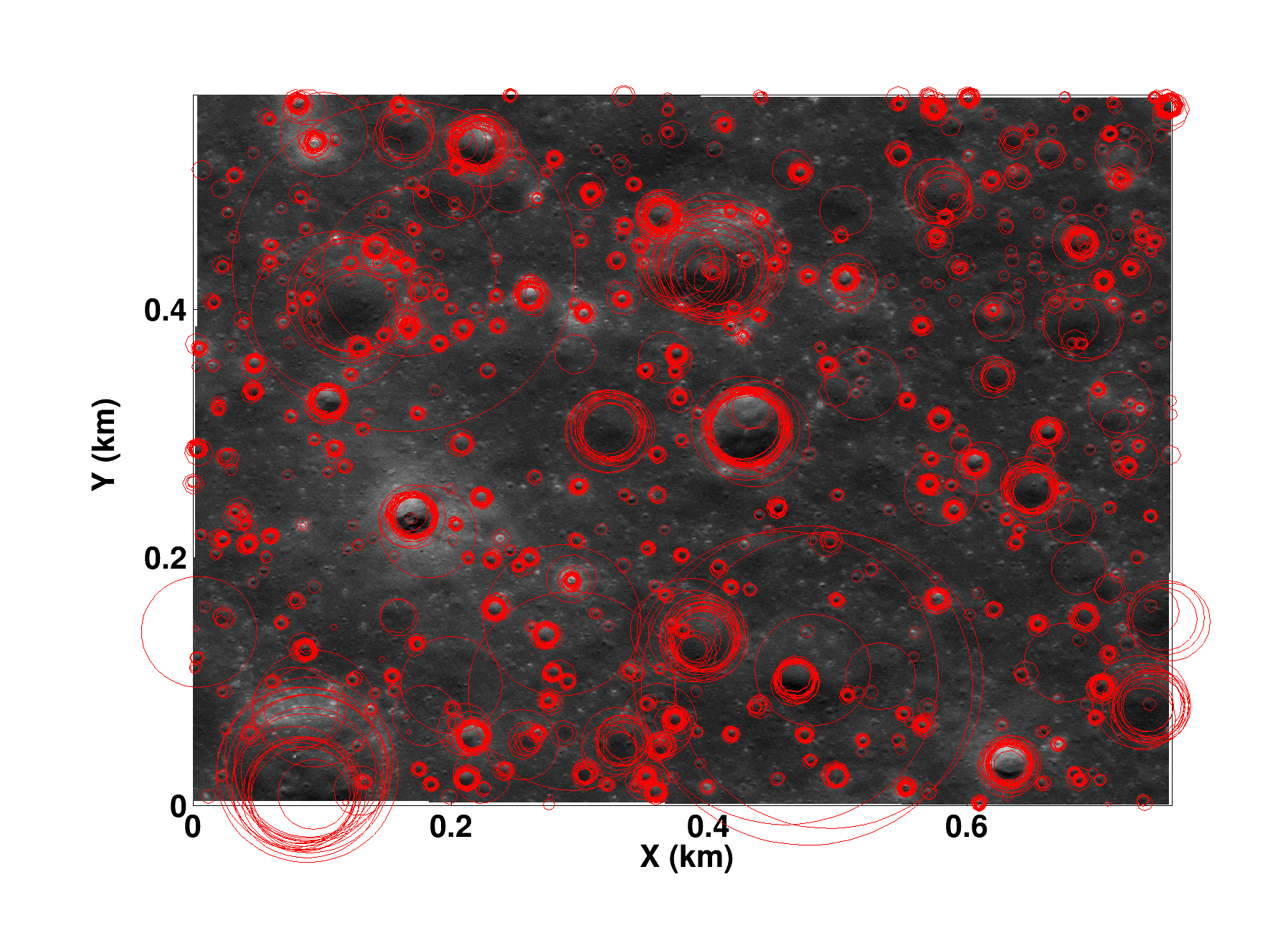}
    }
    
    \subfloat{
        \includegraphics[width=0.99\columnwidth]{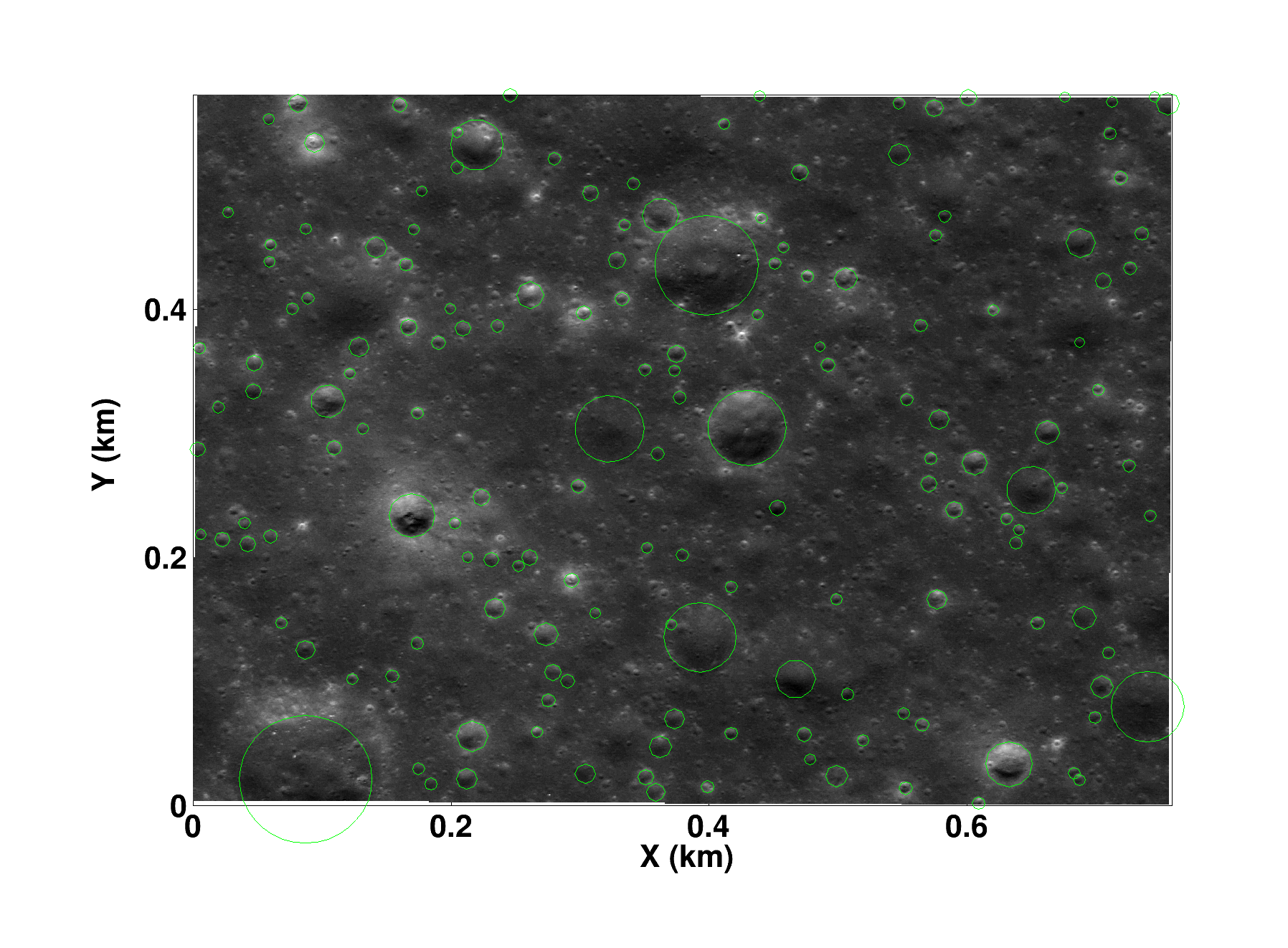}
    }

    \caption{Craters on an original LROC image (top left) are annotated multiple times by volunteers from the public (top right). A clustering algorithm combines the initial markings (with a minimum threshold of N$\geq$9 marks per crater) and assigns a unique marking per crater that is is used to create a final catalog (bottom).}
    \label{fig:craters}
\end{figure*}

\section{Methods}
\label{sec:methods}

In recent years, astronomical surveys and projects have steadily increased their capabilities to produce large quantities of data. As new computational and analysis techniques advance to meet the new challenges that arise, it has also been recognized that involving the general public in astronomical research through community science is also a valuable method for data analysis and verification (\emph{e.g.}, \citet{Raddick_2010, marshall15, bug16}). 

Founded in 2012, CosmoQuest is a virtual research facility that engages the general public in community science projects related to astronomy \citep{robbins12, gugliucci14b, gugliucci14a, gay14}. CosmoQuest's custom web interface has hosted a suite of projects that have not only allowed volunteers to assist in the mapping of craters and other features on the Moon, Mars, Mercury, Vesta, and Bennu, but has also increased the general public's exposure to the various topologies that these worlds exhibit. 

One of the original community projects produced by CosmoQuest is \mm\ \citep{robbins12}. Volunteers gain access to the online interface of \mm\ through CosmoQuest's \mm\ web portal, which also provides tutorials on how to use the interface and educational materials (about the impact of solar incidence on lunar crater identifications, in this case) through blog posts and videos.\footnote{\url{https://cosmoquest.org/x/moon-mappers-tutorials}}$^{,}$\footnote{\url{https://bit.ly/347WQVG}} CosmoQuest's volunteers are also engaged with the project via social media and, in some cases, through more direct communication channels such as {\tt{discord}}, 
{\tt{Twitch}}, and email.  

The \mm\ interface (Fig.~\ref{fig:interface}) is an application specifically designed to facilitate the identification and measurement of craters within the selected LROC NAC images (or \emph{master} images). CosmoQuest's participants have used \mm\ to identify and measure more than 1 million lunar craters \citep{antonenko13}, with a large majority of the crater counts being collected within the first year of \mm\ official launch. All images displayed within the interface are set to 450$\times$450 square pixel sub-images of the master image. Most of the sub-images have the same pixel scale as the {{main}} images; however, some sub-images feature a zoomed-out view to allow for crater identifications at larger diameters. For this analysis, we only use sub-images that are at the original scale. {{ {The images presented to the volunteers have a linear stretch with a range of $\pm$ 0.5 \% (users did not have the ability to change the stretch).}}}

To be an active participant in the project, \emph{i.e.}, for a user's annotations to be recorded, volunteers must be authenticated and have completed the \mm\ tutorial, which covers all essential information and steps to identify craters. The identification of craters is accomplished by selecting the interface's circle annotation tool and drawing circles that match the rim and are centered on craters identified by the user within a given image (Fig.~\ref{fig:interface}).
A minimum threshold diameter of 18 pixels has been imposed to ensure that volunteers are able to visually identify craters with ease (determined through usability testing), restrict the number of crater identifications to a light workload per image, and establish a lower limit, of D $\geq$ 10 m (at the average resolution of the NAC images), on the size of craters to be identified\citep{robbins14}. User-circled annotations that are less than the threshold diameter will appear red and are automatically discarded by the interface. Annotations that are equal to or greater than the threshold are displayed in green and can be modified (resized, re-positioned, or erased). Upon moving to the next image, all identified craters are saved in a MySQL database that can be accessed by project staff for later analysis. We require each 450$\times$450 square pixel sub-image to be annotated by 15 different volunteers in order for it to be considered complete. For calibration and quality assurance purposes, at random intervals the volunteers are shown an image that has already been marked by an expert. The volunteers are given feedback on their scores compared to the experts to encourage quality work. In addition, this calibration process is used to assign confidence intervals to a particular volunteer's work during crater clustering \citep{robbins14}. A 3D clustering algorithm \citep{robbins14} is used to combine the initial markings and assign a unique marking per crater (demanding a minimum of N$\geq$9 marks per crater) that will be saved to produce the final crater catalog (Fig. \ref{fig:craters}).

\begin{figure*}
    \centering
    \subfloat{
        \includegraphics[width=0.99\columnwidth]{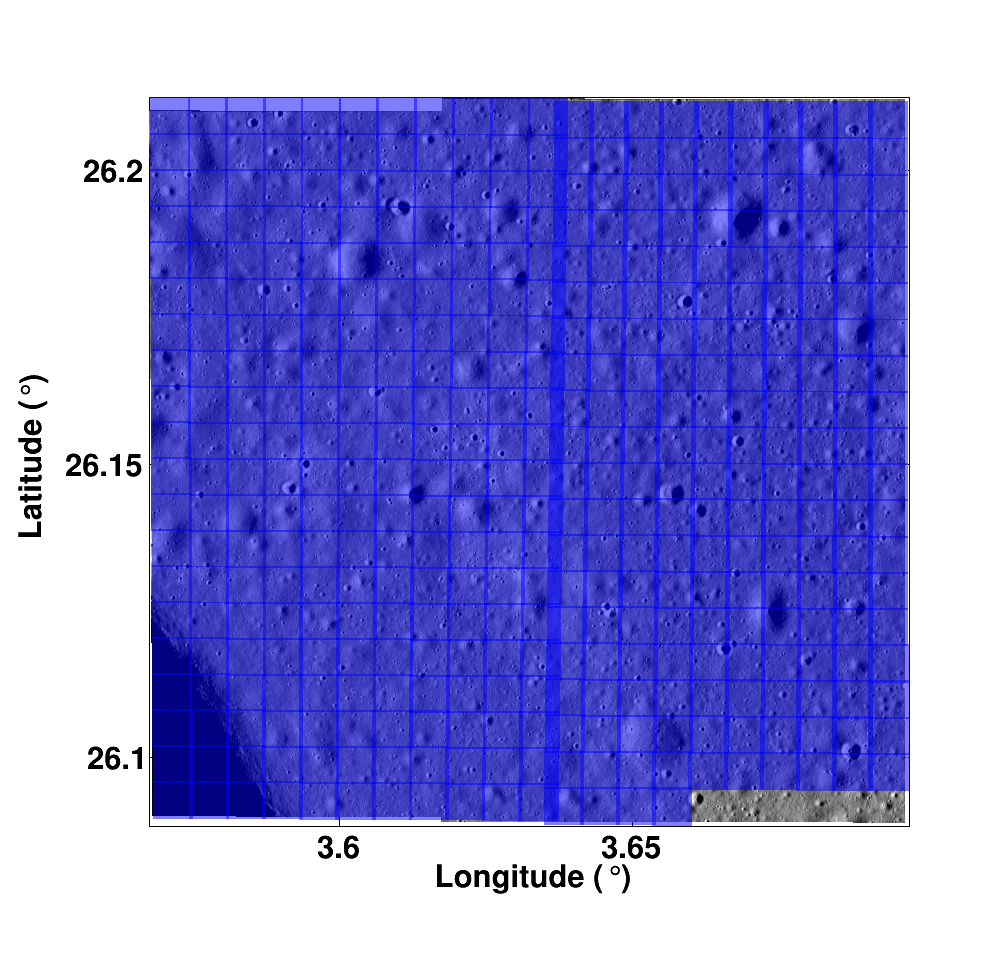}
    }
    \subfloat{
        \includegraphics[width=0.99\columnwidth]{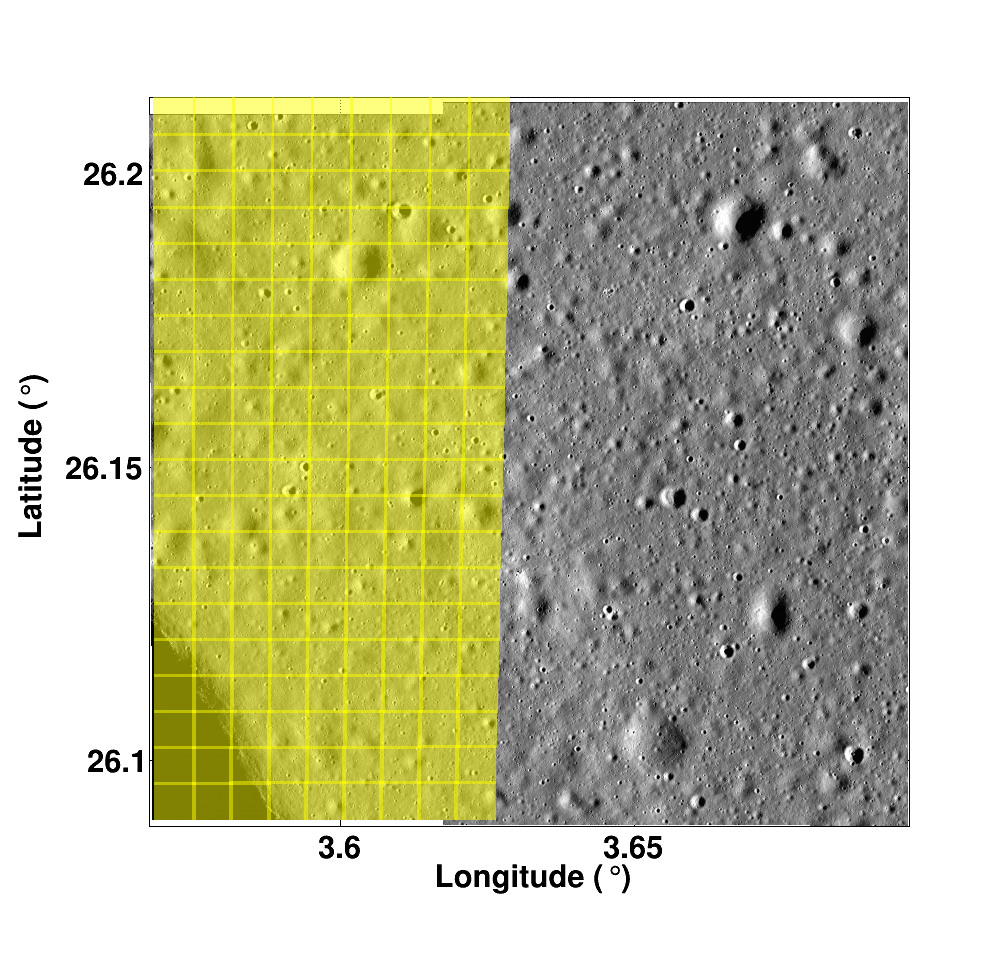}
    }
    
    \subfloat{
        \includegraphics[width=0.99\columnwidth]{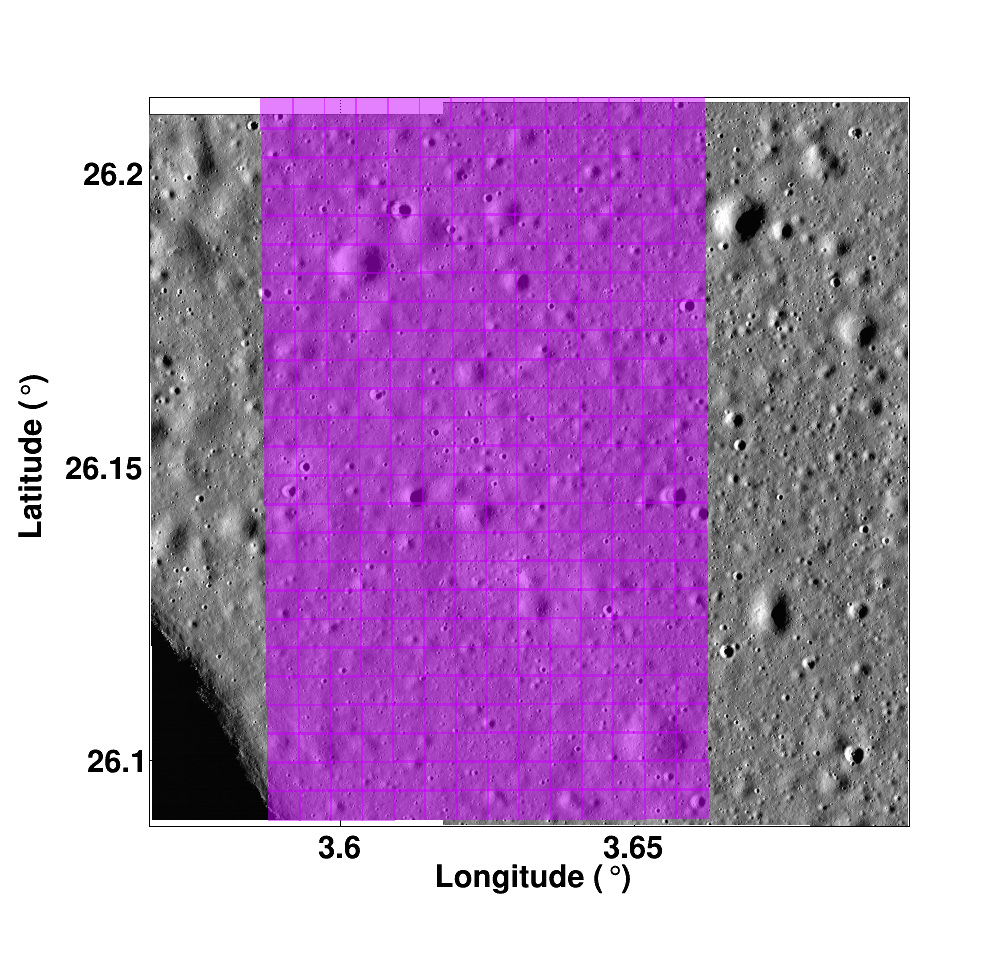}
    }
    \subfloat{
        \includegraphics[width=0.99\columnwidth]{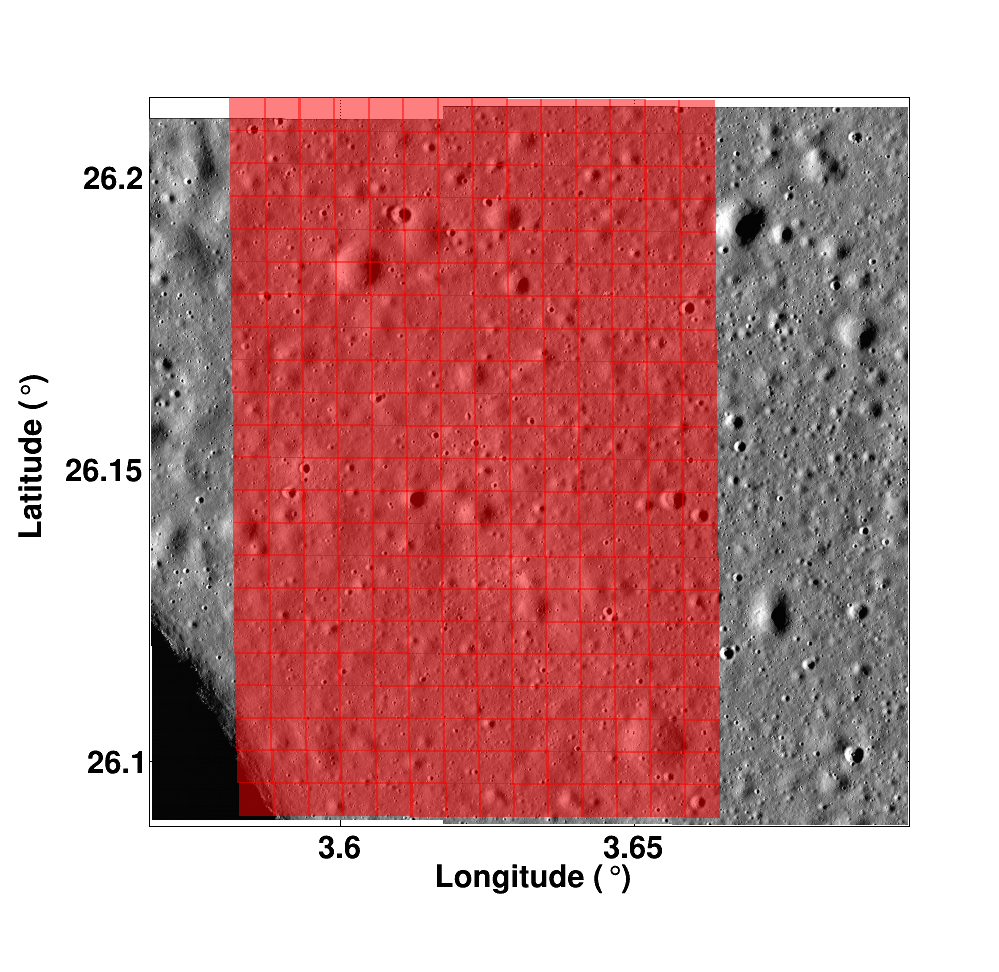}
    }

    \caption{ Image {\tt{M146959973LERE}} at incidence angle of 77.5$^{\circ}$ (background). Each foreground colored transparent image represents the overlap area between {\tt{M146959973LERE}} and one of the other NAC images (from top to bottom and left to right): {\tt{M109215691LERE}}, {\tt{M111578606RE}}, {\tt{M117467833RE}}, and {\tt{M119829425LE}}. The axes are in selenographic coordinates.}
    \label{fig:overlap}
\end{figure*}

\subsection{Preliminary analysis of crater data}
The first year of crater annotations made by \mm\ volunteers on our previous work \citep{antonenko13,grier18} yielded over 1.1 million crater identifications and diameter measurements. Upon using the 3D clustering algorithm, the crowd-sourced crater identifications translated to a catalog of tens of thousands of physical craters. Having amassed such a large database of craters, a statistically significant preliminary investigation into the effects of solar incidence angle on crater identification could be done. At each incidence angle, all craters that had been identified were used to construct cumulative size-frequency distributions (CSFD; histograms of the crater diameters that are summed from large to small diameters). The results showed that when the Sun was higher in the sky, such as $\sim 30^{\circ}$, volunteers had more difficulty finding craters than when the Sun was lower in the sky, such as $\sim 80^{\circ}$. Quantitatively, this discrepancy in the perceived crater population within the same region resulted in approximately four times fewer crater identifications in the crater population enumerated at the lowest solar incidence angle relative to the crater population observed at the highest solar incidence angle. The analysis indicated that the ideal incidence angle for crater identification lies within $\sim 58^{\circ}$ $<$ $i$ $<$ 77$^{\circ}$.\\

The major conclusion drawn from the preliminary analysis agreed with expectations, \emph{i.e.}, that the solar incidence angles closer to zenith makes the identification and measurement of craters difficult. For the results of the preliminary analysis to be valid, two assumptions must be true: 1.) the crater population throughout the \emph{Apollo} 15 landing region is consistent, and 2.) the catalog produced from \mm\ volunteers has high completeness at each solar incidence angle used for the analysis (\emph{i.e.}, at a given angle, volunteers have identified as many craters as a professional crater counter would have identified). Therefore, in this work we have restricted the analysis to regions on the lunar surface that are shared in common between pairs of LROC NAC images taken at different solar incidence angles. We have used the crowd-sourced crater identifications within these regions to derive and compare resulting cratering statistics, as this approach allows us to safely ignore the aforementioned assumptions.

%
%
%

\begin{figure*}[htp]
\centering
\includegraphics[width=.15\textwidth]{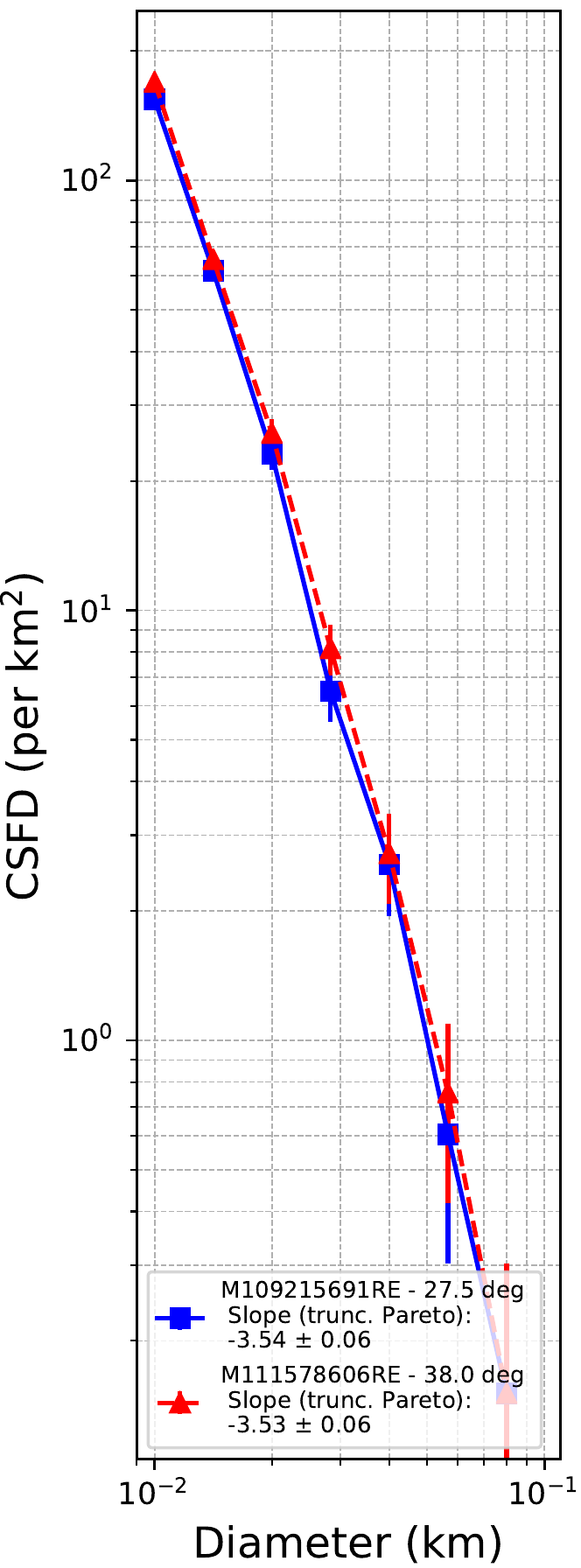}\quad
\includegraphics[width=.15\textwidth]{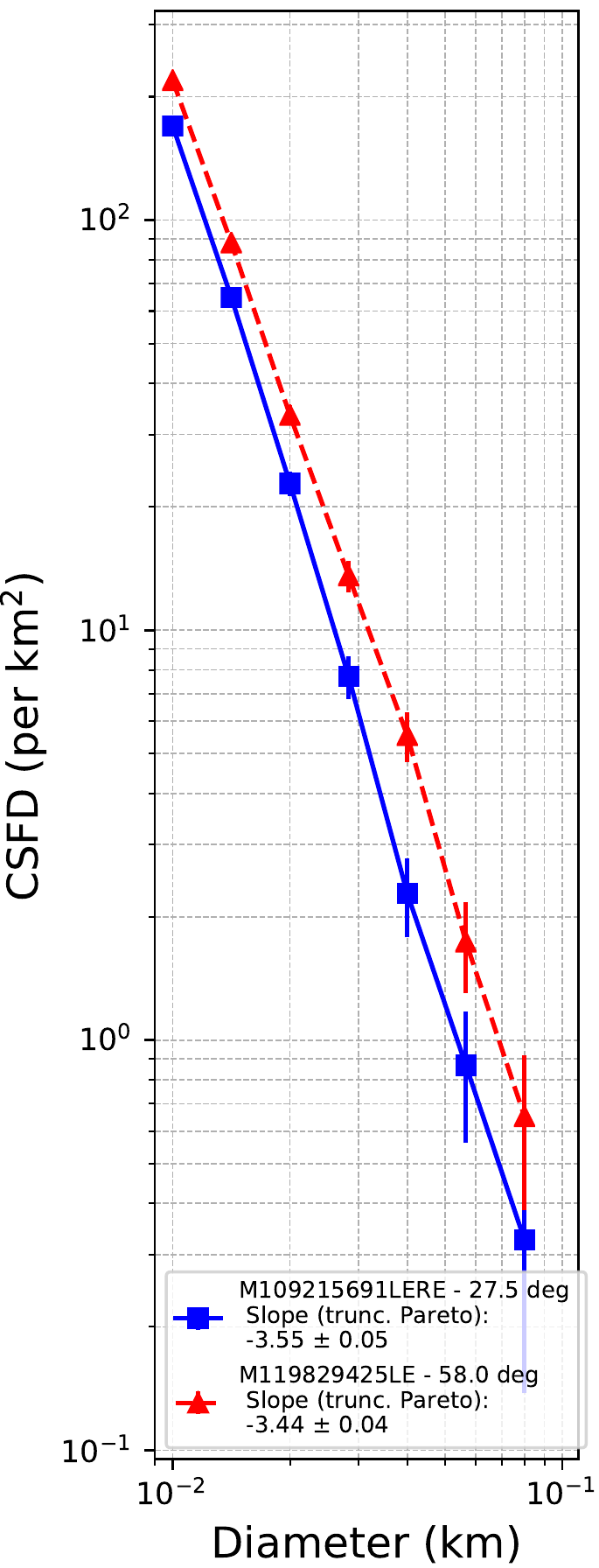}\quad
\includegraphics[width=.15\textwidth]{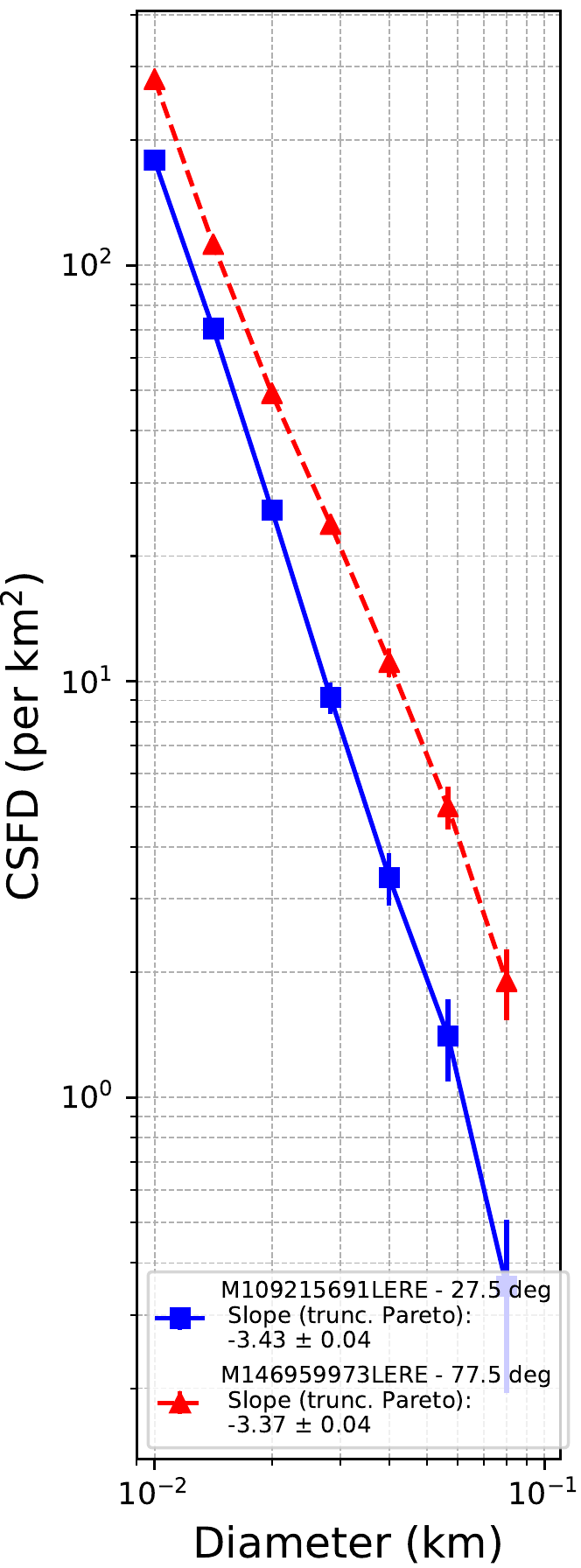}
\includegraphics[width=.15\textwidth]{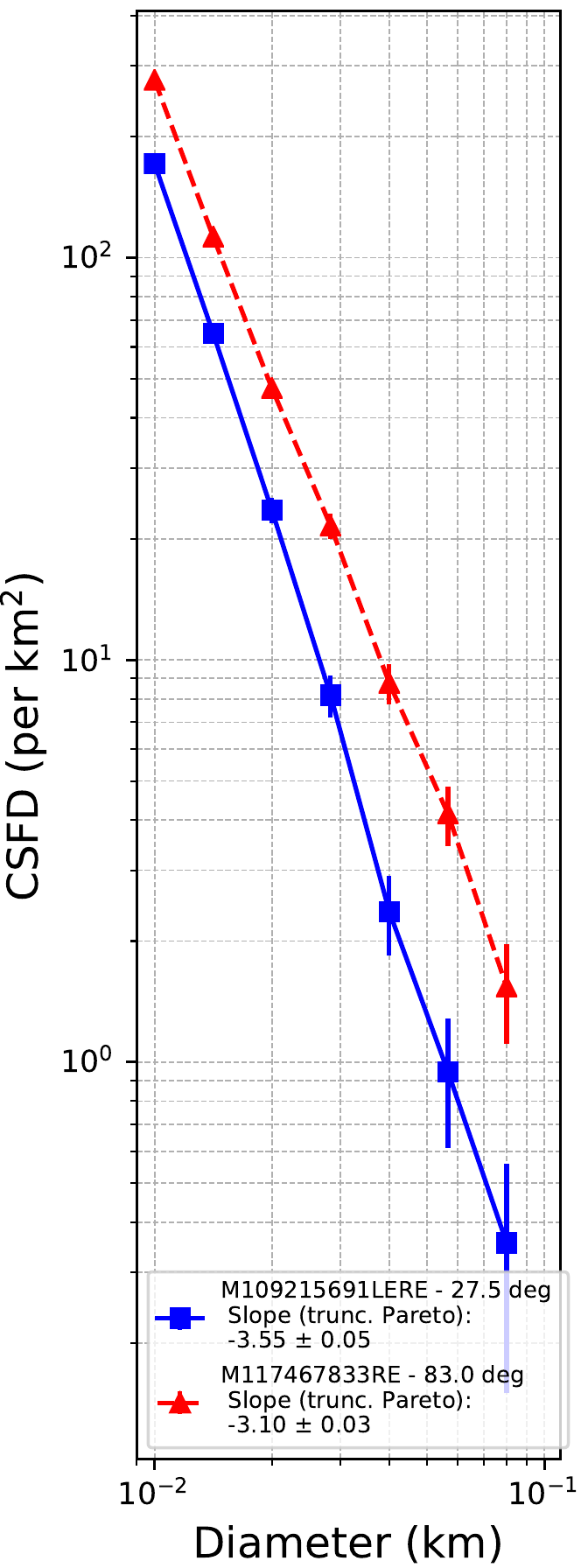}\quad
\includegraphics[width=.15\textwidth]{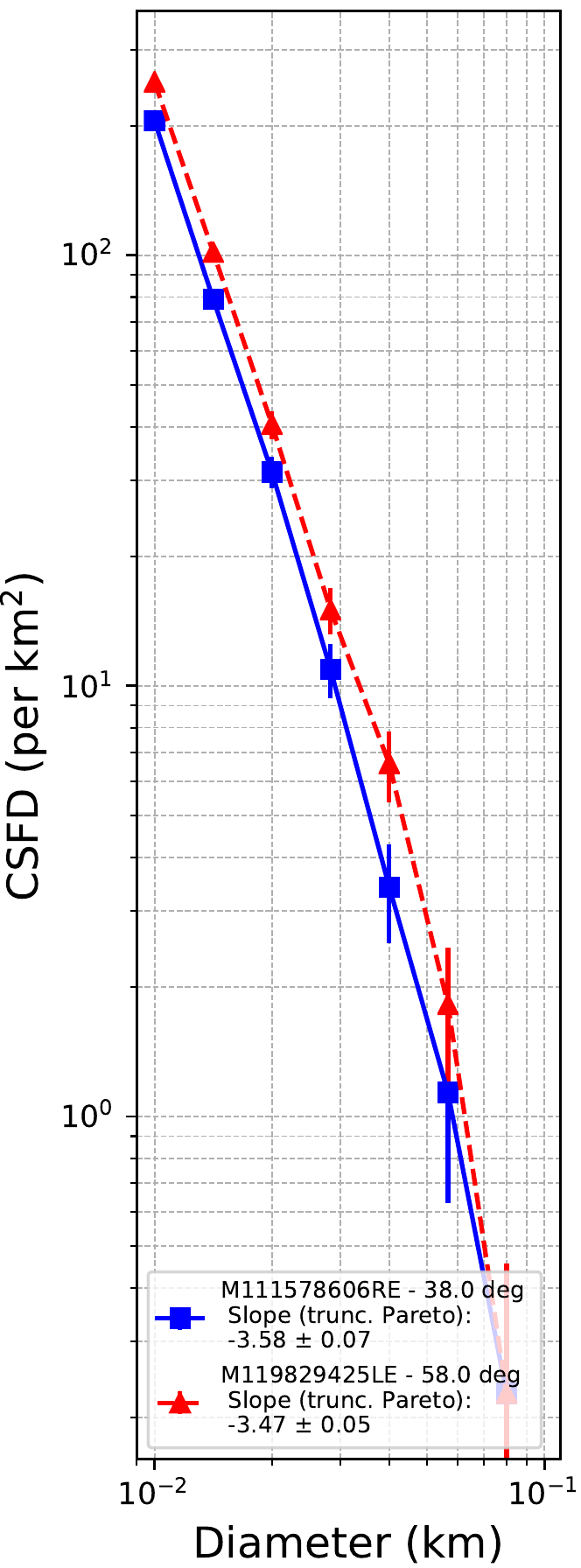}
\medskip

\includegraphics[width=.15\textwidth]{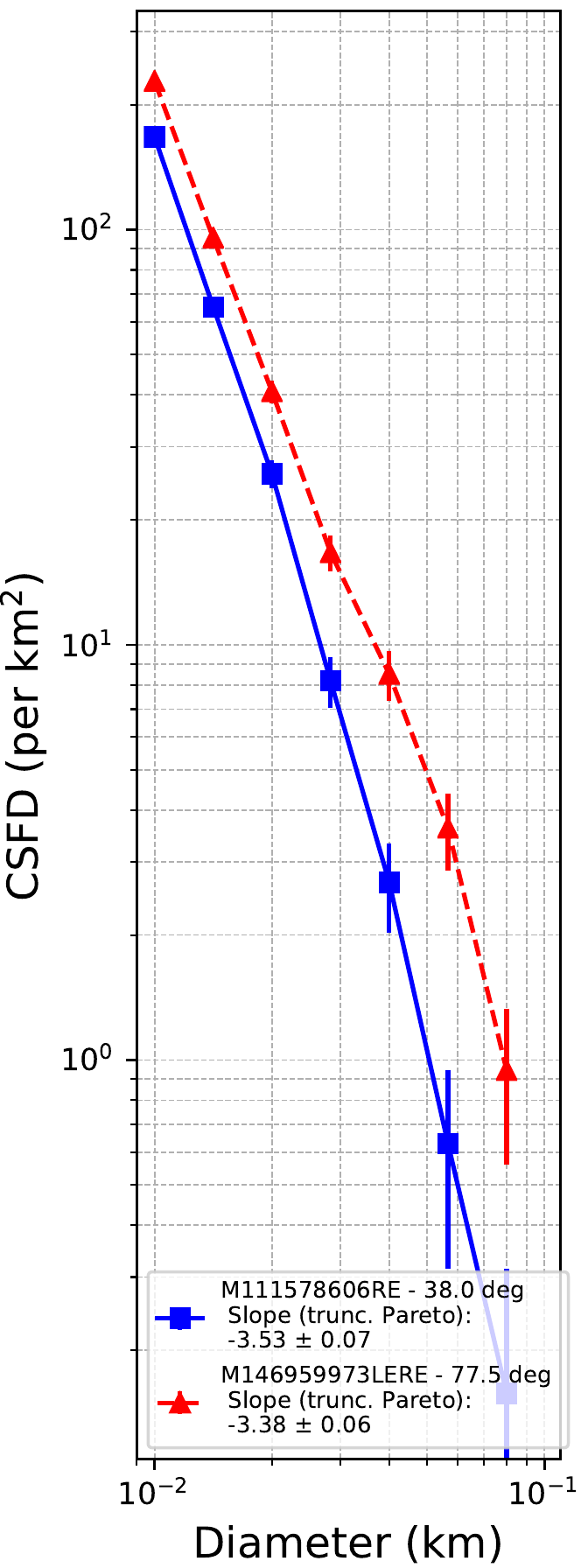}\quad
\includegraphics[width=.15\textwidth]{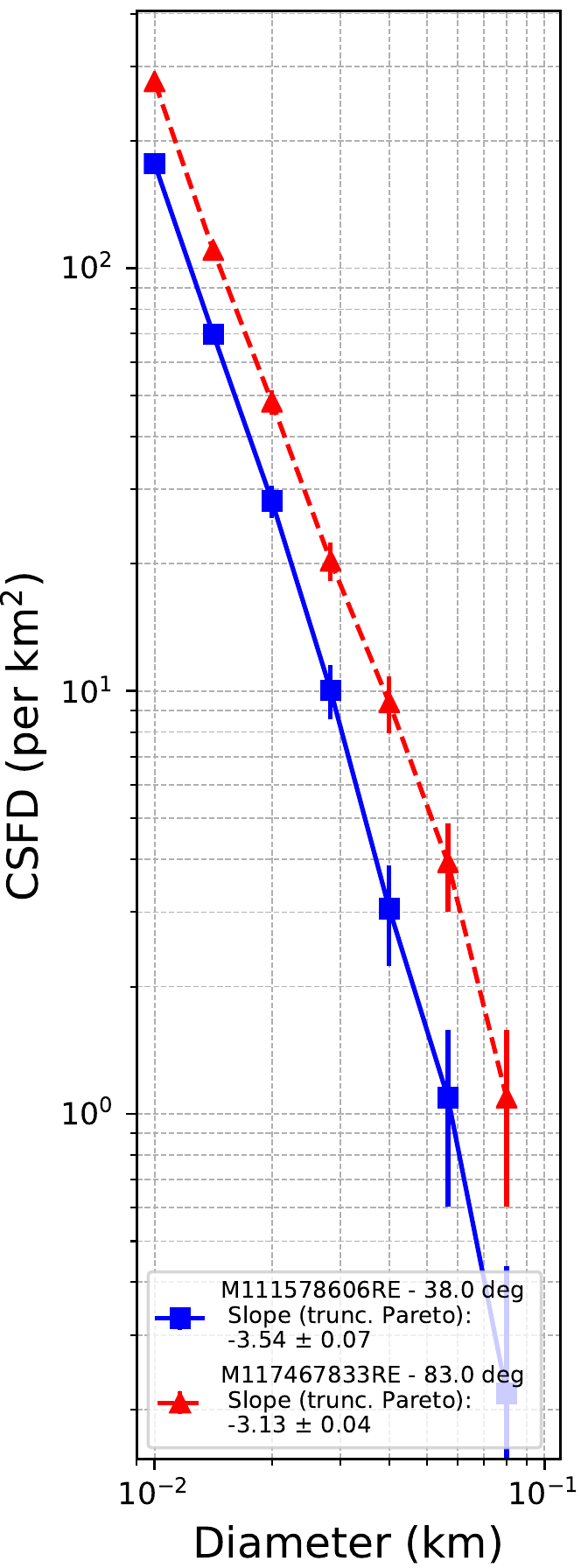}\quad
\includegraphics[width=.15\textwidth]{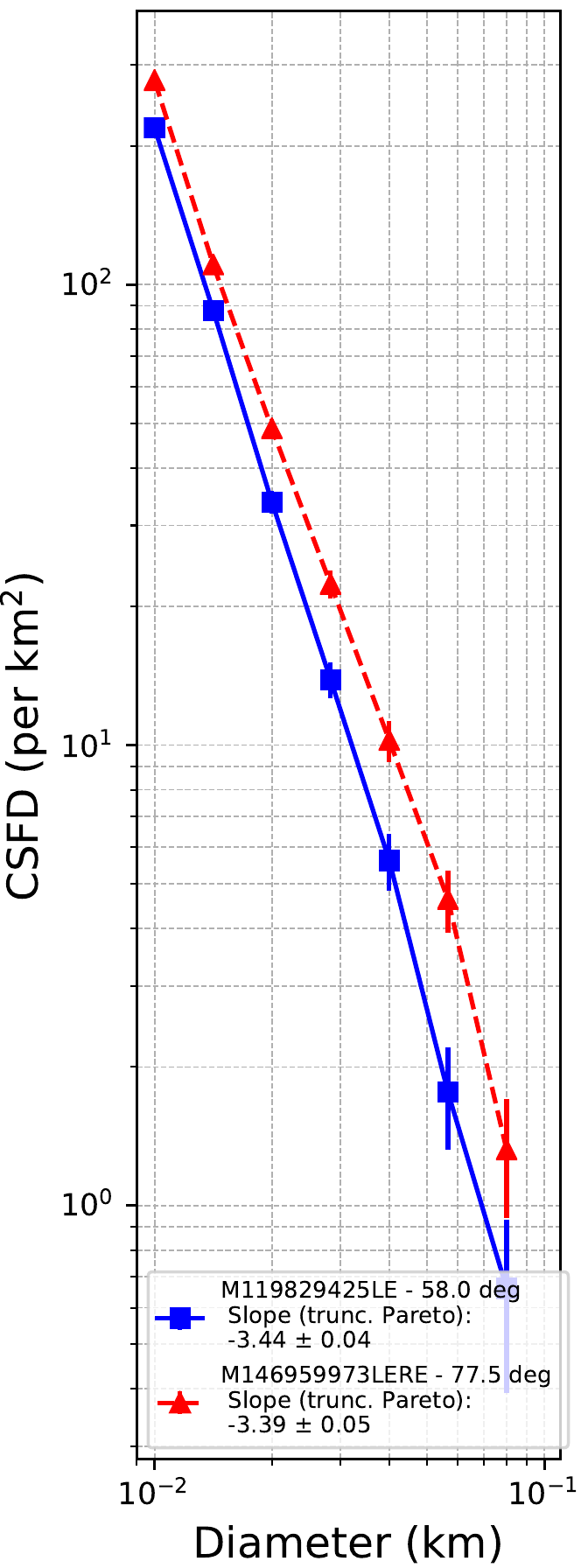}
\includegraphics[width=.15\textwidth]{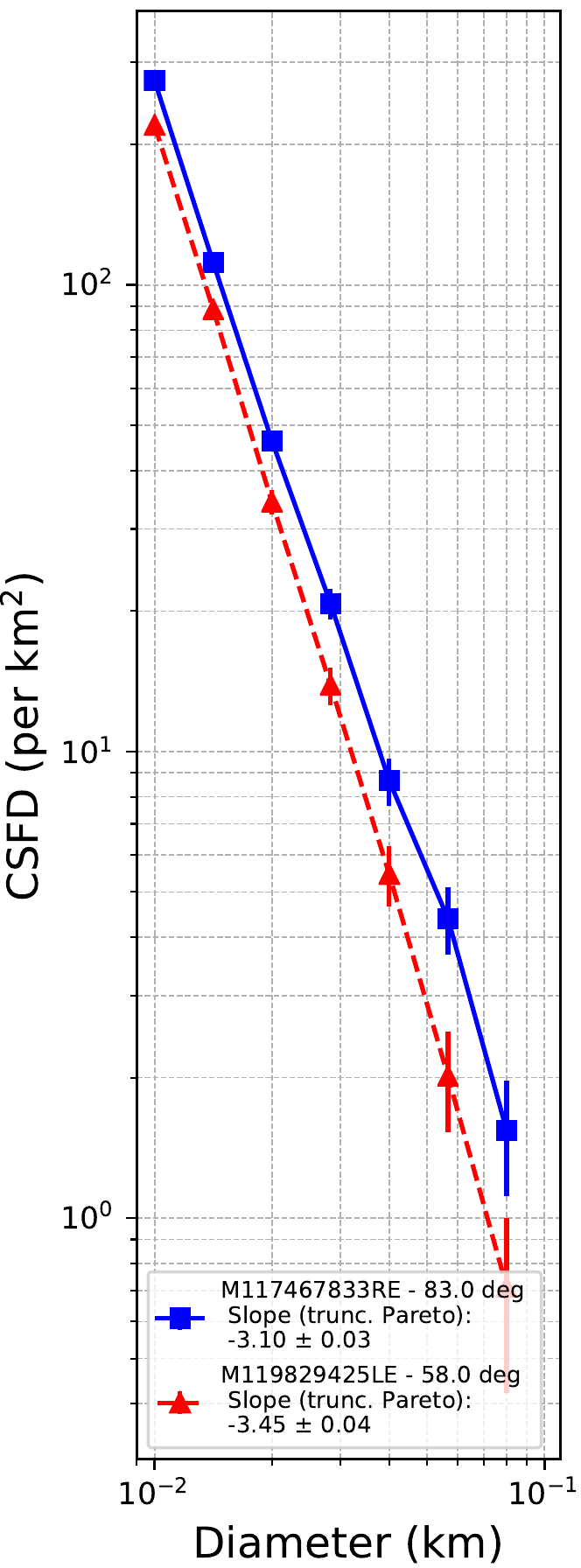}\quad
\includegraphics[width=.15\textwidth]{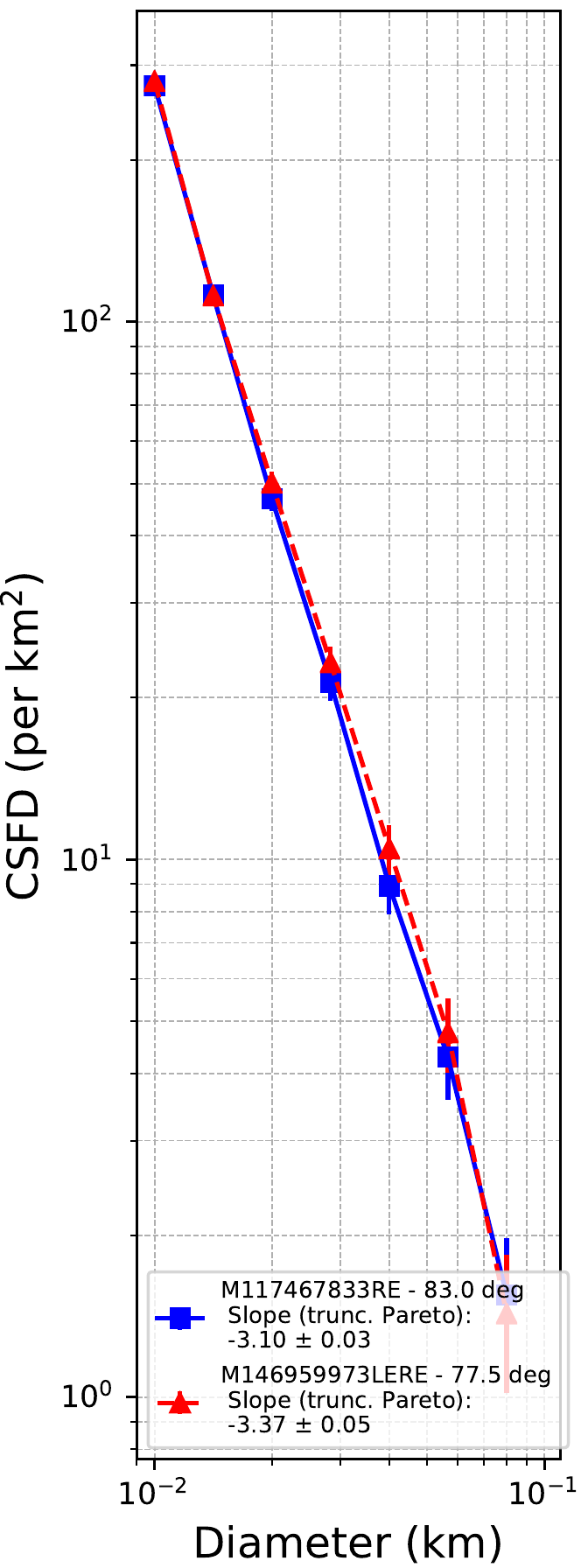}

\caption{Cumulative size-frequency distributions (CSFD) plots. Each panel depicts the difference in the CSFD and for craters identified within the same area imaged at two different incidence angles. Colors and plotting style indicate which distribution was derived from the larger incidence angle (dashed and triangles) and smaller incidence angle (solid and squares). {The slope value for a fit to a truncated Pareto distribution of the unbinned diameter data is also shown in the legend for each incidence angle.}}
\label{fig:csdf}
\end{figure*}

\begin{figure*}[htp]
\centering
\includegraphics[width=.15\textwidth]{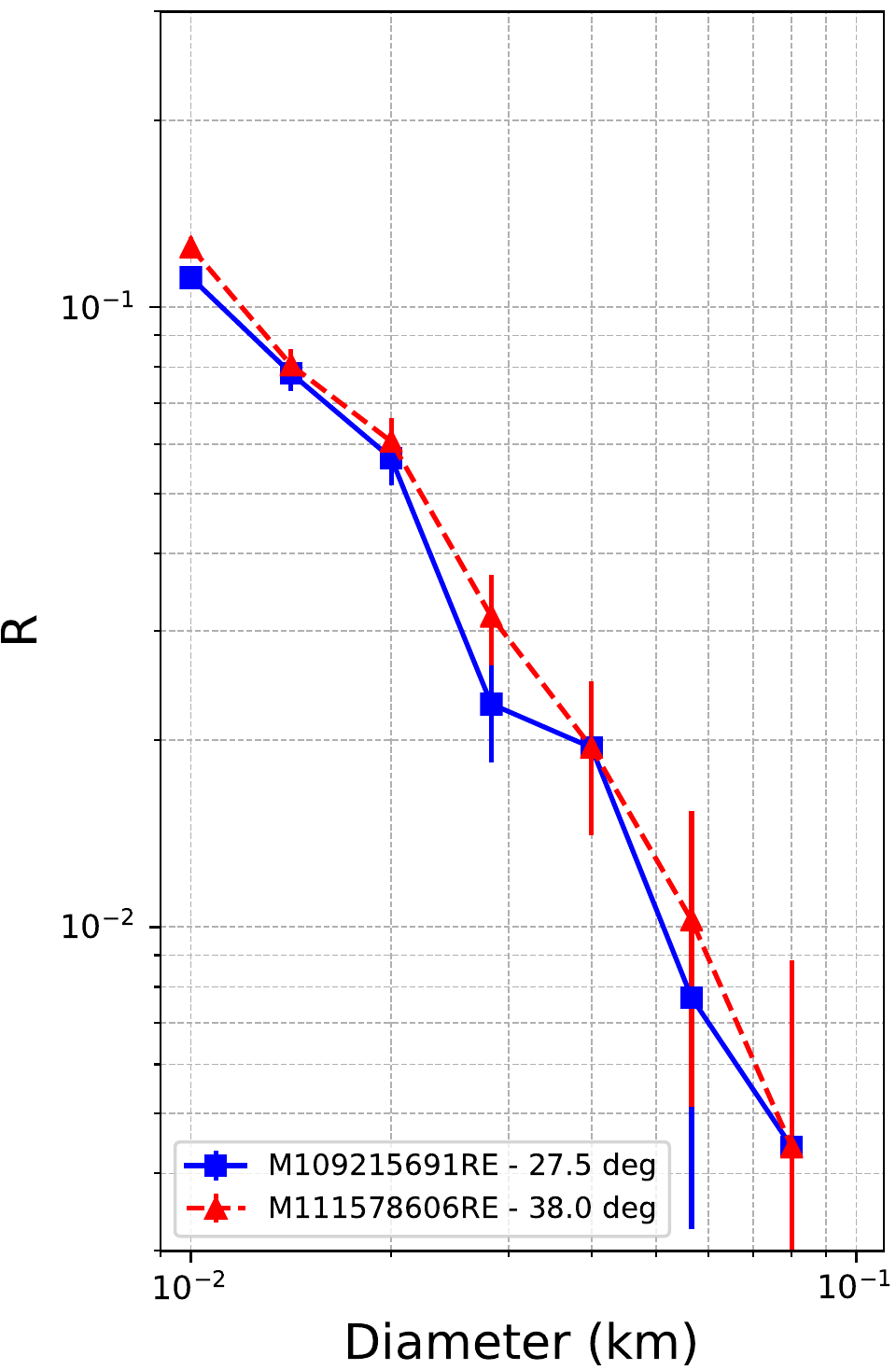}\quad
\includegraphics[width=.15\textwidth]{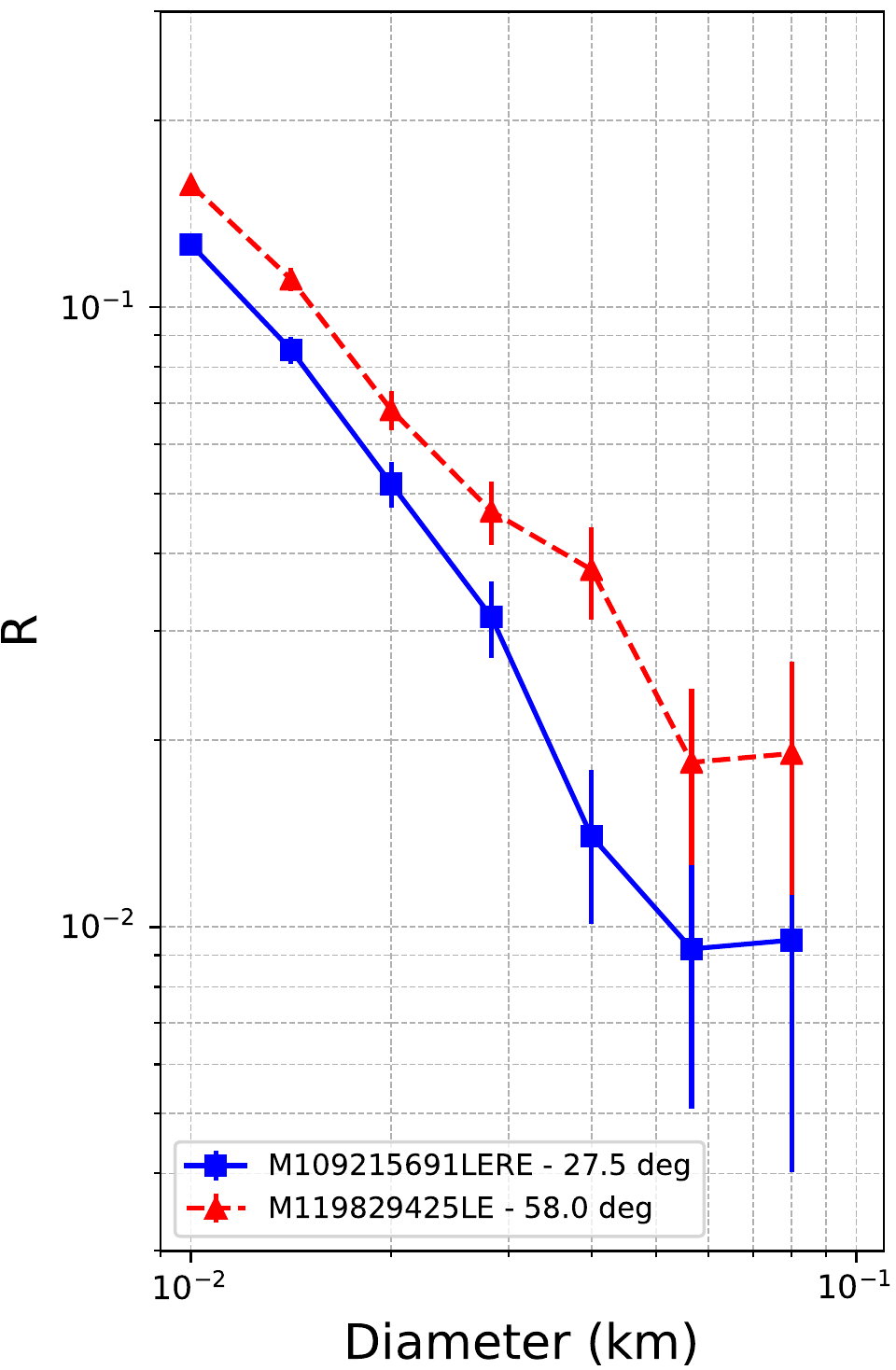}\quad
\includegraphics[width=.15\textwidth]{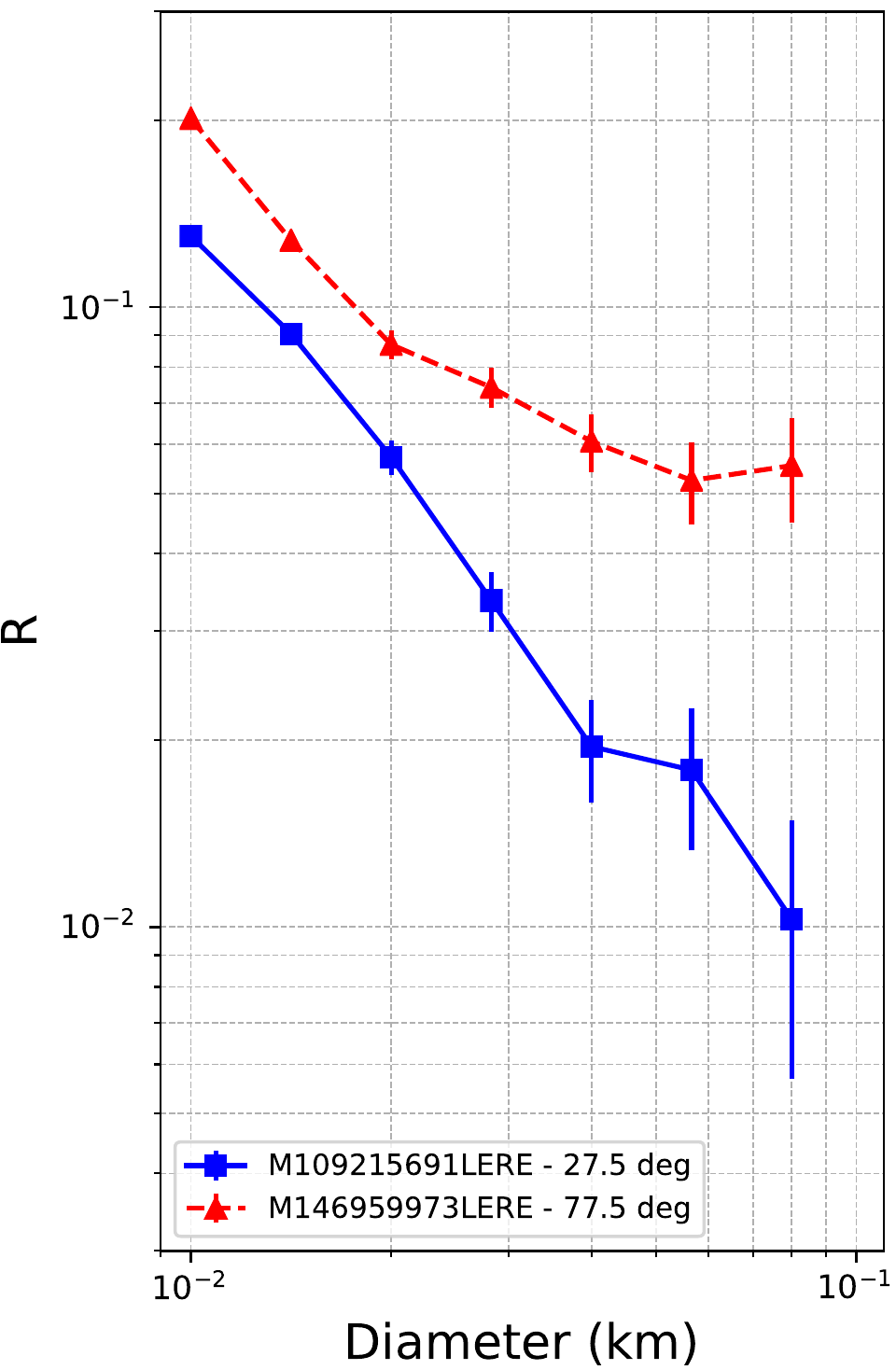}
\includegraphics[width=.15\textwidth]{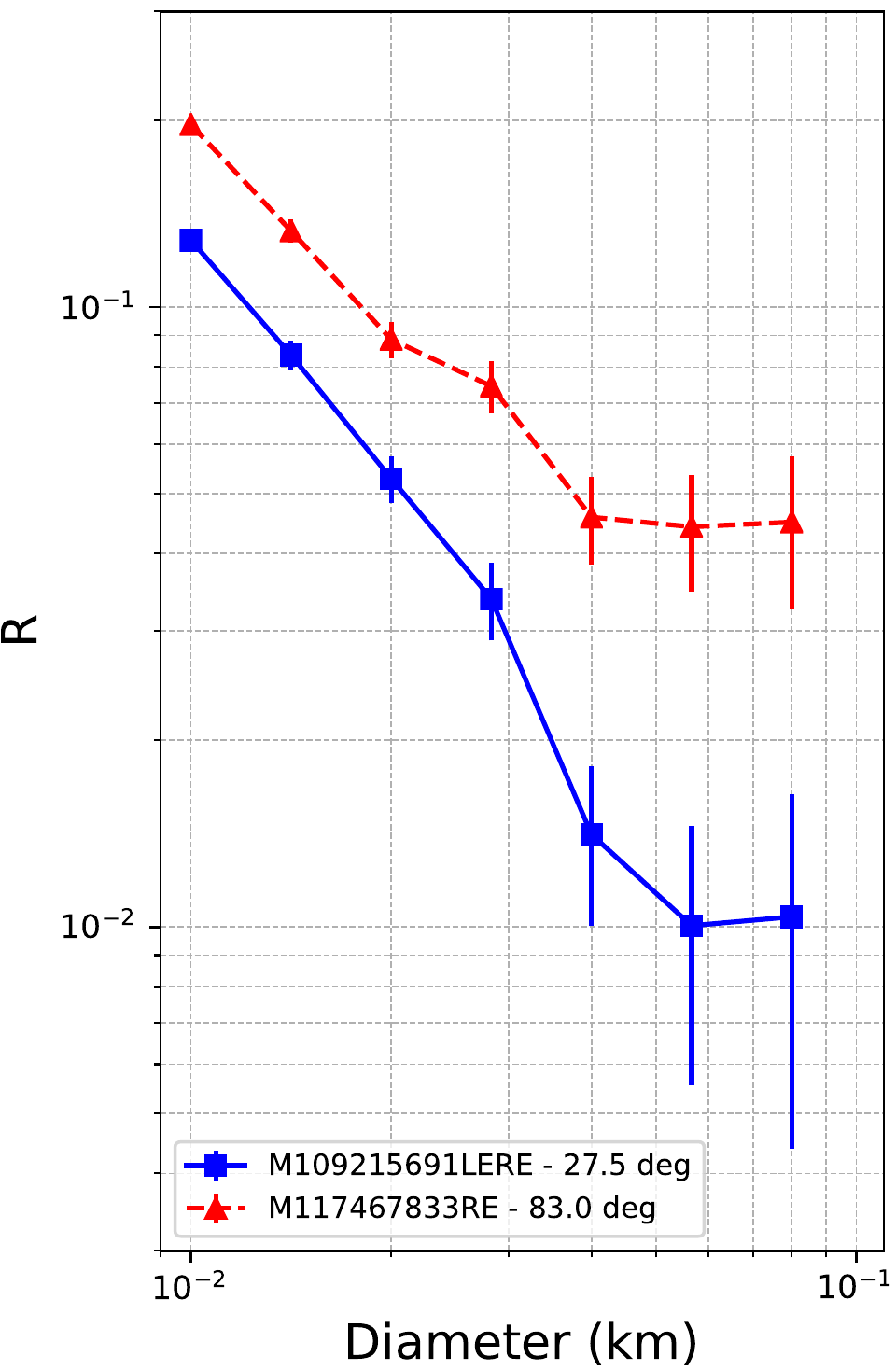}\quad
\includegraphics[width=.15\textwidth]{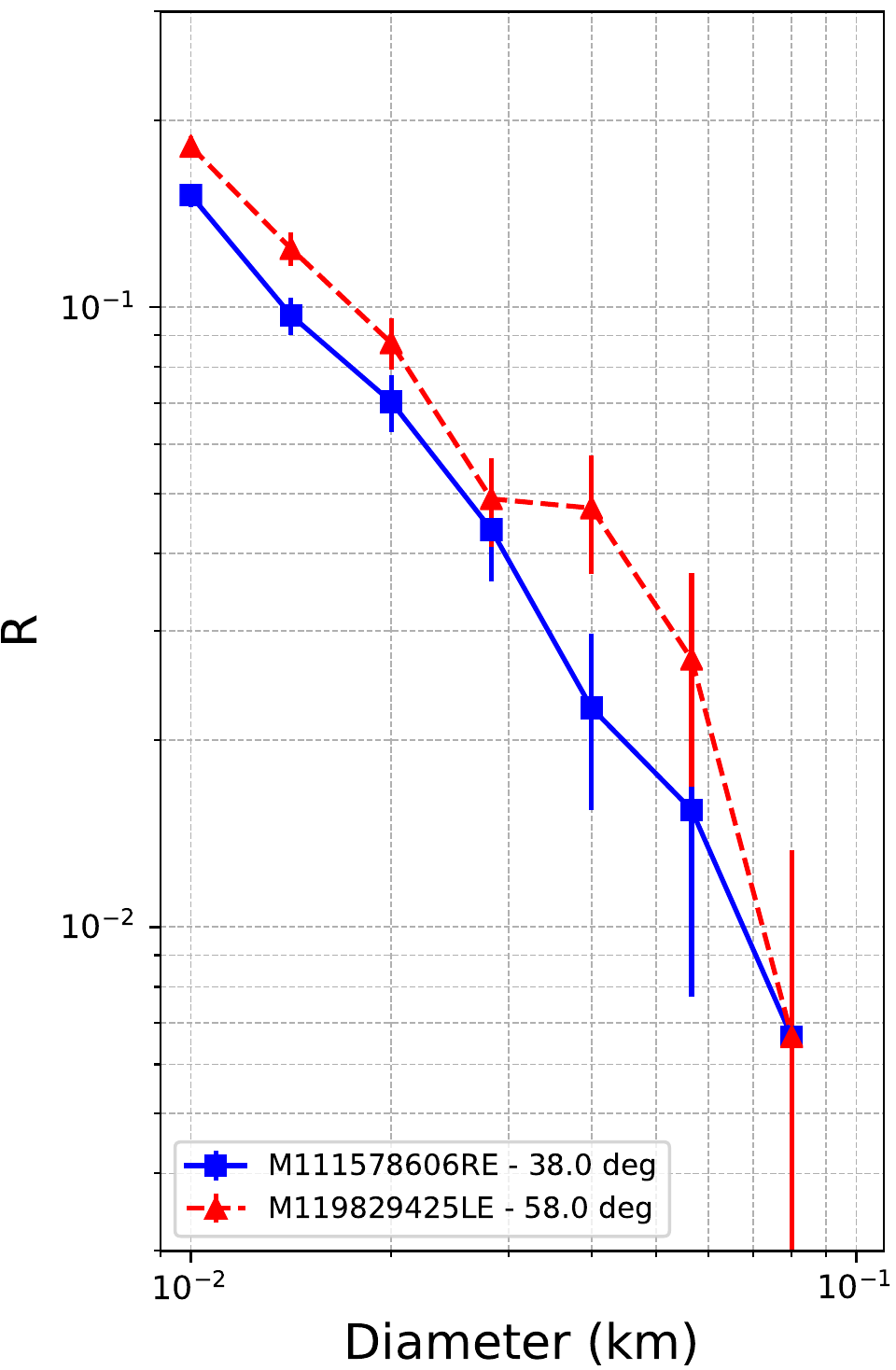}
\medskip

\includegraphics[width=.15\textwidth]{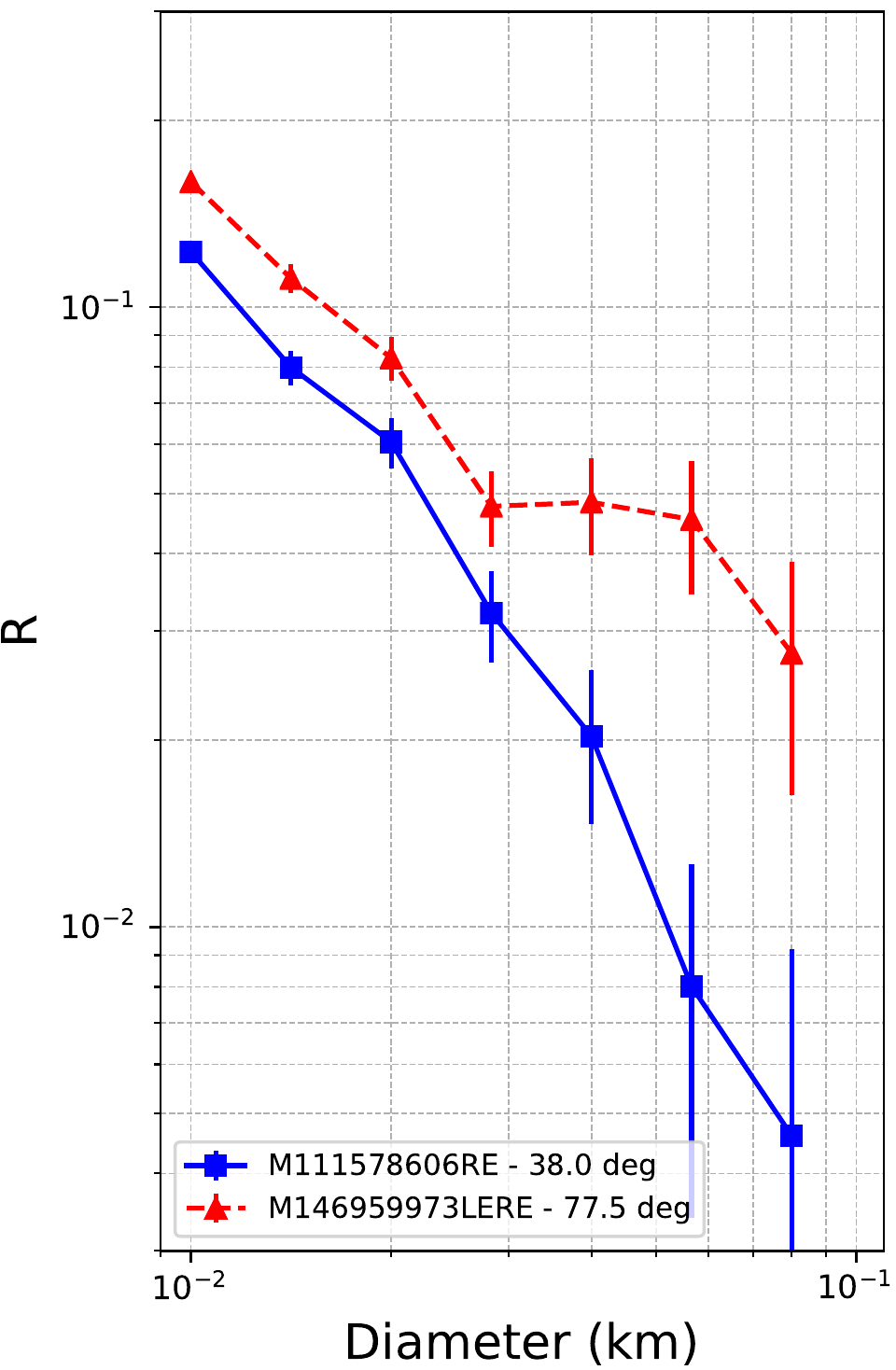}\quad
\includegraphics[width=.15\textwidth]{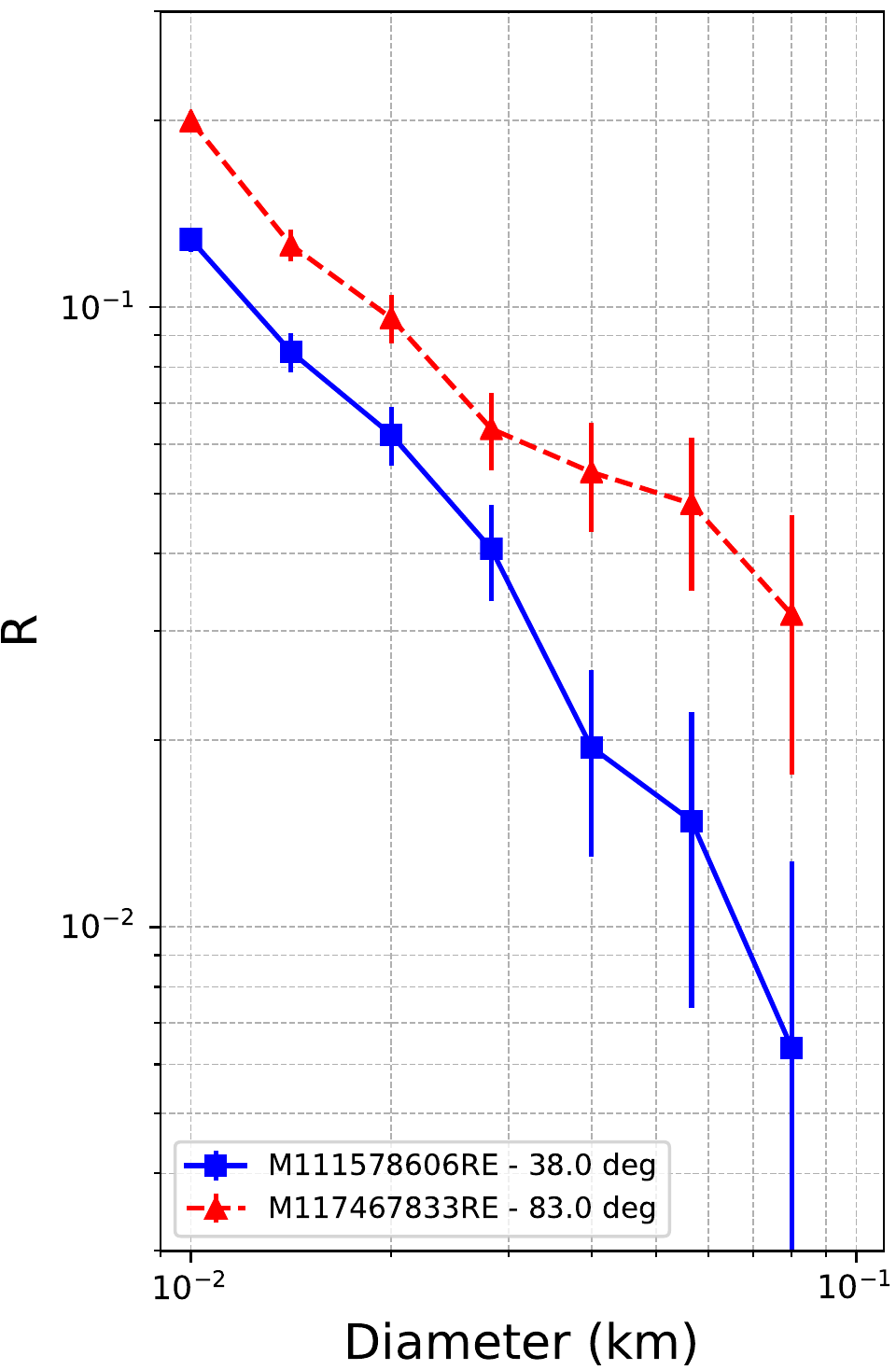}\quad
\includegraphics[width=.15\textwidth]{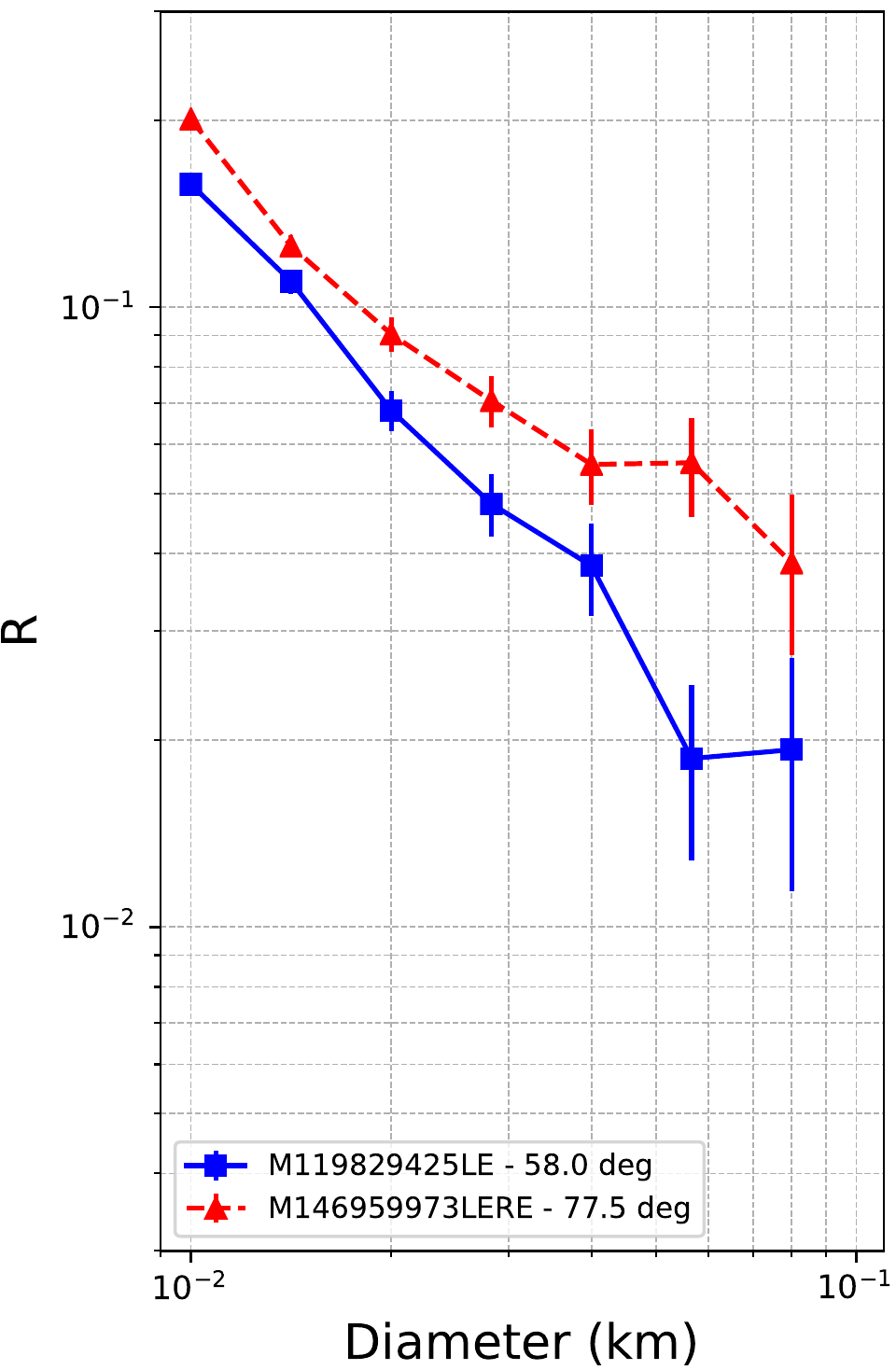}
\includegraphics[width=.15\textwidth]{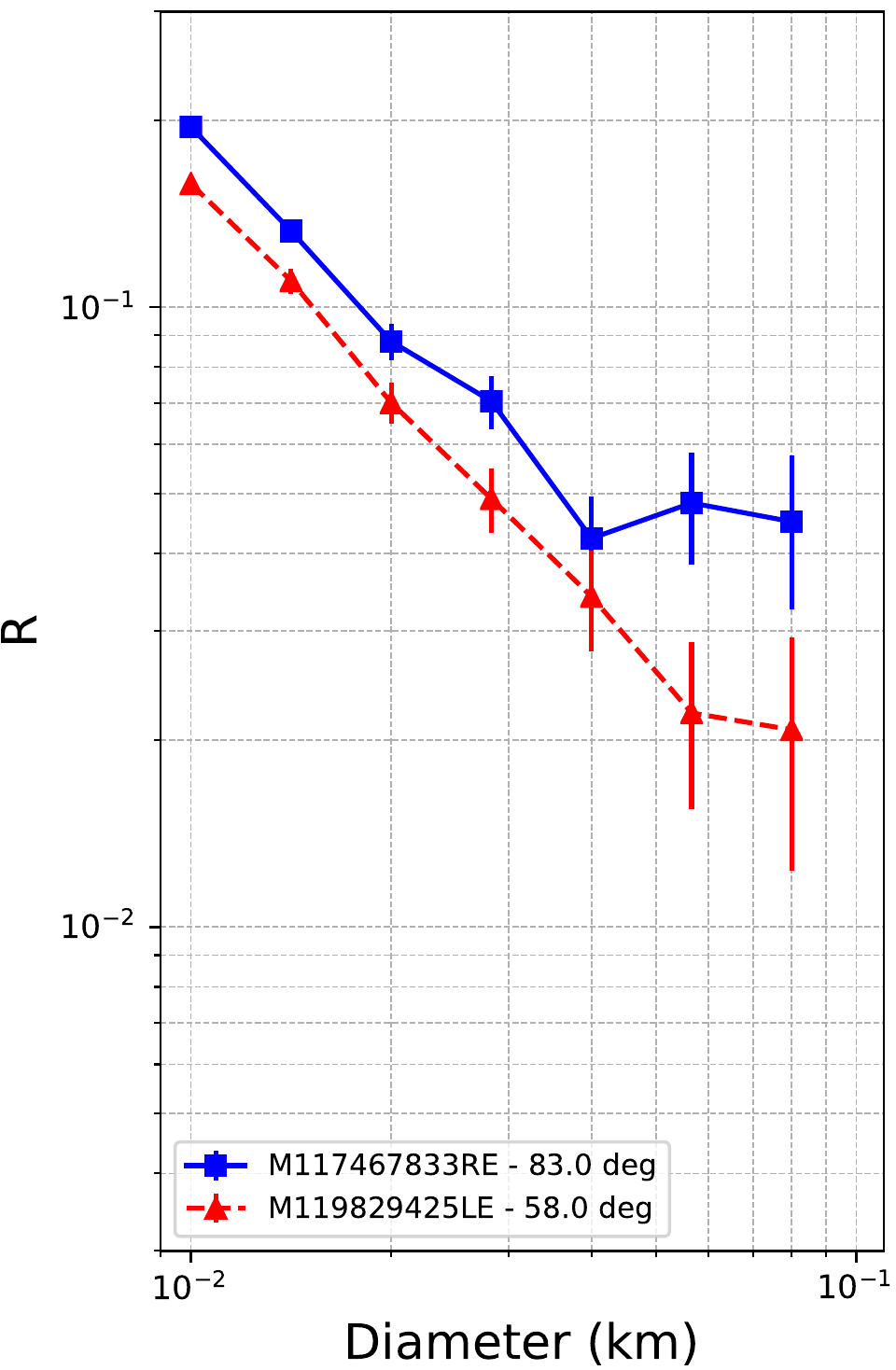}\quad
\includegraphics[width=.15\textwidth]{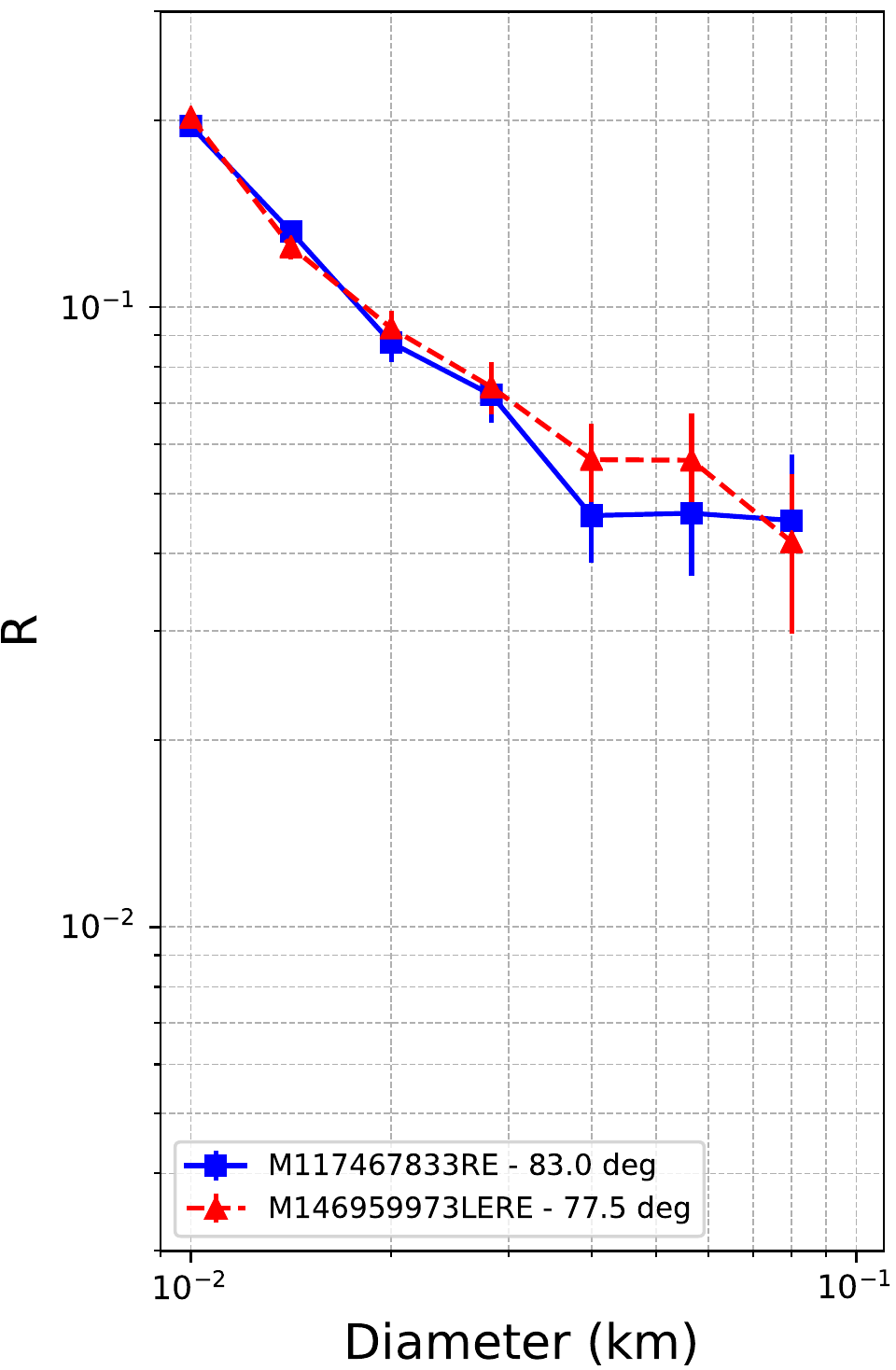}

\caption{``R" plots. Each panel depicts the difference in the ``R" and for craters identified within the same area imaged at two different incidence angles. Colors and plotting style indicate which distribution was derived from the larger incidence angle (dashed and triangles) and smaller incidence angle (solid and squares).}
\label{fig:rplots}
\end{figure*}

\subsection{Crater analysis in overlapping regions}
While the master images generally cover the same area, the overlap between images at different incidence angles is not complete. To ensure that the non-overlapping areas would not bias comparisons made between different sets of calculated cratering statistics derived from images at different incidence angles, we first identified the overlapping region between a given set of master images at two different incidence angles (Fig.~\ref{fig:overlap}).\\

The overlap is determined at the 450$\times$450 square pixel scale so as to only include sub-images that had been examined by \mm\ volunteers. To find the pairs of overlapping sub-images, we began by first determining the corner coordinates of each sub-image. This was accomplished by making use of the fact that all of the master images' corner coordinates have been defined in selenographic coordinates. The relationship between a given master image's four corner pixel coordinates and corresponding selenographic coordinates allows for a unique mapping between the two coordinate systems through the calculation of the projective transformation matrix.
We calculated the transformation matrix for each master image and then used the transformation matrices to convert the sub-images' four corner pixel coordinates to selenographic coordinates. With the selenographic coordinates of the corners of each sub-image known, we then employed the ray-casting algorithm, also referred to as the crossing number algorithm, to determine which sub-images overlap by checking if any of the vertices of one sub-image were within the bounding region of another sub-image. The Southerland-Hodgman algorithm \citep{sutherland74} was then used to determine the coordinates of the boundary of the overlap region between pairs of sub-images that were determined to have overlap.\\

Visual inspection of the master images indicated that the overlap region between different pairs of master images can differ in either size or geographical location. This fact was accentuated after computationally identifying the overlap regions between the master images. As we wish for our statistical inferences to be based on crater data obtained within the same region imaged at two incidence angles, we elected to calculate statistical results for each pair of master images. This task required us to first identify all user annotations contained within a given overlap region, and record those annotations to produce a catalog of user annotations. Given that a single crater can have multiple user annotations, we utilized a modified density-based spacial clustering of applications with noise (DBSCAN) algorithm---the aforementioned 3D clustering algorithm---to compile a crater catalog of single craters associated with clusters of user annotations in the user annotations catalog.\\

Lastly, we constructed cumulative size-frequency distributions and relative size-frequency distributions (``R" plots) for each pair of incidence angles. We bin and construct the distributions following the conventions in \citet{cratergroup79}. 

\section{Results}
\label{sec:results}
{Figs.~\ref{fig:csdf} and \ref{fig:rplots}} show the resulting CSFDs and ``R" plots for different pairs of incidence angles. Volunteer identifications and measurements of craters are consistent within uncertainties for incidence angle differences $\leq$ 10.5$^{\circ}$ as shown by cumulative and relative SFDs. We begin to see noticeable deviations between cumulative and relative SFDs where the incidence angle difference is $\geq$ 19.5$^{\circ}$. Specifically, the deviations are small at smaller crater diameters and gradually increase in magnitude with increasing crater diameter.

This trend becomes more pronounced with increased difference in incidence angle. For comparisons where the difference in incidence angle is $\geq$ 50$^{\circ}$ the CSFD's and ``R" plots appear to be fairly distinct distributions in favor of more craters being identified at the higher incidence angle.

{The relative comparison between low and high incidence angles becomes more informative under the assumption that the high-angle images ($ \geq 77^{\circ}$) provide a better crater-mapping surface and are therefore more representative of the actual crater population in this image. Lower angles ($\leq 58^{\circ}$) are presumptively more susceptible to recognition challenges, \emph{i.e.,} they are more susceptible to observational or perceptual variables rather than actual surface conditions, and therefore rendering the images analyzed under these conditions potentially less optimal for science investigations.}

{Following \citet{robbins18}, we also
fit the unbinned crater size data 
to a truncated Pareto distribution.\footnote{\citet{robbins18} argue that  this maximum likelihood method is more reliable than the more traditional least-squares method that fits a power-law.} The slopes of the distribution are reported in the legend of Fig.~\ref{fig:csdf} for each incidence angle and field. We note that the fitted slopes are $< -3$, which would indicate the areas are still in production. However, it is known that the \emph{Apollo 15} sites are in equilibrium, with cumulative slopes in the derived data of approximately $-2$ or shallower (\emph{e.g.}, \citet{hirabashi17} and the results from CosmoQuest's previous study in \citet{robbins14}). The results in this study show a general agreement with our previous results in \cite{robbins14} at high angles within the errors bars, however, the results at lower angles ($< 58^{\circ}$) are not in good agreement at larger diameters, which could contribute to steepen the slopes in this case. A factor that could affect identification of larger craters at low solar incidence angle is the roughness of lunar topography (\emph{e.g.}, \citet{kreslavsky13} have analyzed the roughness of the lunar surfaces at large range of scales). At high solar incidence angles (lunar evening/morning), the general background roughness of the lunar surface is comparable to small craters, and therefore more of them might be missed under these conditions. On the other hand, at low solar incidence angles (near lunar noon) the almost absence of shadows and more presence of reflective differences is on the scale of smaller craters and the eye of the non-subject matter experts latches onto those features, preferentially finding them. However, in large craters, these prominent topographic cues are not present, and instead, the eyes of the volunteers are drawn to features \emph{within} the craters that are at the same scale as smaller craters elsewhere.}

\section{Conclusions}
\label{sec:discussion}

The results we have obtained from the data confirm that \mm\ volunteers identified more lunar craters when the Sun is lower on the horizon (high incidence angle) on overlapping regions. We conclude that for a given area on the lunar surface, imaged at two different incidence angles, one should expect that the overall number of craters identified at the higher incidence angle will be greater than the overall number of craters identified at the lower incidence angle. Furthermore, a larger difference in incidence angle should generally result in a larger difference in the overall number of craters identified at the two levels of surface illumination. In particular, we find that there is practically no difference when the sun is very low on the horizon ($77.5 ^{\circ}$ vs $83.0 ^{\circ}$) or high in the sky ($27.5 ^{\circ}$ vs $38.0^{\circ}$), but we found progressively fewer craters were found as sun angle decreases between $\approx 40^{\circ}$ and $\approx 75^{\circ}$.  Moving from a $\approx 75^{\circ}$ -- $80^{\circ}$ incidence angle to a mere $30^{\circ}$–-$40^{\circ}$ cuts crater detections by a maximum of $\approx 3.5x$ (see \emph{e.g.,} the cumulative and relative counts from Table 3 in the appendix).  We find that this difference decreases for the smallest craters, possibly because $\approx$ $10$--$20$ pixel craters are easier to interpret as depressions regardless of sun angle than the larger ones.

{{Compared to our previous the results in \citet{robbins14}, the results of this present study show an agreement consistent with the error bars, except for smaller scales, where the data presented in this study tends to be higher, and at larger diameters for smaller incidence angles. We point out, however, that  our study used a larger area compared to \citet{robbins14} (approximately $2.5x$ larger area; see tables in the appendix), which could also contribute to the mismatch at these scales. In the case that there were a systematic offset towards smaller biases in the CosmoQuest's \mm\ data and under the assumption that is is uniform across different incidence angles, our study would still be valid given that its main point is to analyze incidence angle differences via crowd sourcing. }} 

{{In order to complement the observational analysis presented here, an additional venue of study would be the study of the topographical characteristics of the non-detected craters at low incidence angles, compared to those detected, via complementary data such as the \emph{Apollo 15} landing site Digital Terrain Models for areas that overlap. In addition, further work could include the analysis of the magnitude of incidence angle effects at size ranges larger than those considered in this study ($ \geq 100$ m, \emph{e.g.}, $200$ m -- $2$ km) where the population is not expected to be fully degraded and the impact of incidence angle should therefore diminish. This type of study would also allow to gauge the interplay between observational and topographical/geological factors.}}

Our results validate and verify what many professional crater counters' analyses have been conveying for a few decades now. For a cratered region of a planetary surface that has been imaged at two considerably different incidence angles, one expects to be able to identify more craters in the image captured at the higher incidence angle. {This effect is qualitatively well understood in terms of the brightness variations caused by variable illumination of different slopes:} at lower incidence angles with respect to the normal (when the sun is high overhead), craters are more difficult to distinguish, given that in these circumstances albedo is more prominent and topographic features are more difficult to identify \citep{wilhelms87}. {{We highlight that the results of this analysis were generated with contributions from volunteers of the general public. This study underscores the value in having the public as science collaborators by confirming previous work and helping refine confidence levels, and fundamentally demonstrates the viability of public participation in authentic scientific investigations in future studies}}

\section{Acknowledgments}
The authors would like to thank the many CosmoQuest volunteers whose hard work and dedication made this study possible. We thank the anonymous referees for their suggestions, which have improved the clarity and readability of the paper. Funding for CosmoQuest's \mm\ project and research staff was provided through NASA cooperative agreements NNX16AC68A and NNX17AD20A, grants NNX09AD34G and NNX12AB92G, and contracts through the Lunar Reconnaissance Orbiter education and public outreach office. This publication was made possible through additional support from the NASA SSERVI VORTICES project, via CAN NNA14AB02A. {{This research has made use of NASA's Astrophysics Data System Bibliographic Service, and of the following software: {{\tt NumPy}} \citep{2011CSE....13b..22V}, {{\tt Matplotlib}} \citep{Hunter:2007}}, and {{\tt SciPy}} \citep{jones01}}.
\newpage

\bibliographystyle{mnras}
\bibliography{incidence_angle}

\begin{thebibliography}{}
\makeatletter
\relax
\def\mn@urlcharsother{\let\do\@makeother \do\$\do\&\do\#\do\^\do\_\do\%\do\~}
\def\mn@doi{\begingroup\mn@urlcharsother \@ifnextchar [ {\mn@doi@}
  {\mn@doi@[]}}
\def\mn@doi@[#1]#2{\def\@tempa{#1}\ifx\@tempa\@empty \href
  {http://dx.doi.org/#2} {doi:#2}\else \href {http://dx.doi.org/#2} {#1}\fi
  \endgroup}
\def\mn@eprint#1#2{\mn@eprint@#1:#2::\@nil}
\def\mn@eprint@arXiv#1{\href {http://arxiv.org/abs/#1} {{\tt arXiv:#1}}}
\def\mn@eprint@dblp#1{\href {http://dblp.uni-trier.de/rec/bibtex/#1.xml}
  {dblp:#1}}
\def\mn@eprint@#1:#2:#3:#4\@nil{\def\@tempa {#1}\def\@tempb {#2}\def\@tempc
  {#3}\ifx \@tempc \@empty \let \@tempc \@tempb \let \@tempb \@tempa \fi \ifx
  \@tempb \@empty \def\@tempb {arXiv}\fi \@ifundefined
  {mn@eprint@\@tempb}{\@tempb:\@tempc}{\expandafter \expandafter \csname
  mn@eprint@\@tempb\endcsname \expandafter{\@tempc}}}

\bibitem[\protect\citeauthoryear{{Antonenko}, {Robbins}, {Gay}, {Lehan}  \&
  {Moore}}{{Antonenko} et~al.}{2013}]{antonenko13}
{Antonenko} I.,  {Robbins} S.~J.,  {Gay} P.~L.,  {Lehan} C.,   {Moore} J.,
  2013, in Lunar and Planetary Science Conference. p.~2705

\bibitem[\protect\citeauthoryear{{Bell}}{{Bell}}{2020}]{bell20}
{Bell} S.~W.,  2020, \mn@doi [Journal of Geophysical Research (Planets)]
  {10.1029/2020JE006392}, \href
  {https://ui.adsabs.harvard.edu/abs/2020JGRE..12506392B} {125, e06392}

\bibitem[\protect\citeauthoryear{{Bugiolacchi}, {Bamford}, {Tar}, {Thacker},
  {Crawford}, {Joy}, {Grindrod}  \& {Lintott}}{{Bugiolacchi}
  et~al.}{2016}]{bug16}
{Bugiolacchi} R.,  {Bamford} S.,  {Tar} P.,  {Thacker} N.,  {Crawford} I.~A.,
  {Joy} K.~H.,  {Grindrod} P.~M.,   {Lintott} C.,  2016, \mn@doi [\icarus]
  {10.1016/j.icarus.2016.01.021}, \href
  {https://ui.adsabs.harvard.edu/abs/2016Icar..271...30B} {271, 30}

\bibitem[\protect\citeauthoryear{{Chin} et~al.,}{{Chin} et~al.}{2006}]{chin06}
{Chin} G.,  et~al., 2006, in {Mackwell} S.,  {Stansbery} E.,  eds, 37th Annual
  Lunar and Planetary Science Conference. Lunar and Planetary Science
  Conference.
p.~1949

\bibitem[\protect\citeauthoryear{{Christensen}, {Engle}, {Anwar},
  {Dickenshied}, {Noss}, {Gorelick}  \& {Weiss-Malik}}{{Christensen}
  et~al.}{2009}]{christensen09}
{Christensen} P.~R.,  {Engle} E.,  {Anwar} S.,  {Dickenshied} S.,  {Noss} D.,
  {Gorelick} N.,   {Weiss-Malik} M.,  2009, AGU Fall Meeting Abstracts, \href
  {http://adsabs.harvard.edu/abs/2009AGUFMIN22A..06C} {pp IN22A--06}

\bibitem[\protect\citeauthoryear{{Crater Analysis Techniques Working Group}
  et~al.,}{{Crater Analysis Techniques Working Group}
  et~al.}{1979}]{cratergroup79}
{Crater Analysis Techniques Working Group} et~al., 1979, \mn@doi [\icarus]
  {10.1016/0019-1035(79)90009-5}, \href
  {http://adsabs.harvard.edu/abs/1979Icar...37..467C} {37, 467}

\bibitem[\protect\citeauthoryear{{Elliot}, Huang, Minton  \& M}{{Elliot}
  et~al.}{2018}]{elliot18}
{Elliot} R.~J.,  Huang Y.,  Minton D.~A.,   M F.~A.,  2018, Icarus, 312, 231

\bibitem[\protect\citeauthoryear{Farley et~al.,}{Farley
  et~al.}{2014}]{farley14}
Farley K.~A.,  et~al., 2014, \mn@doi [Science] {10.1126/science.1247166}, 343

\bibitem[\protect\citeauthoryear{{Gault}}{{Gault}}{1970}]{gault70}
{Gault} D.~E.,  1970, \mn@doi [Radio Science] {10.1029/RS005i002p00273}, \href
  {http://adsabs.harvard.edu/abs/1970RaSc....5..273G} {5, 273}

\bibitem[\protect\citeauthoryear{{Gay}, {Lehan}, {Moore}, {Bracey}  \&
  {Gugliucci}}{{Gay} et~al.}{2014}]{gay14}
{Gay} P.~L.,  {Lehan} C.,  {Moore} J.,  {Bracey} G.,   {Gugliucci} N.,  2014,
  in Lunar and Planetary Science Conference. Lunar and Planetary Science
  Conference.
p.~2927

\bibitem[\protect\citeauthoryear{{Grier}, {Richardson}, {Gay}, {Robbins},
  {Lehan}, {Antonenko}  \& {CosmoQuest Team}}{{Grier} et~al.}{2018}]{grier18}
{Grier} J.~A.,  {Richardson} M.,  {Gay} P.,  {Robbins} S.,  {Lehan} C.,
  {Antonenko} I.,   {CosmoQuest Team} 2018, in Lunar and Planetary Science
  Conference. p.~2479

\bibitem[\protect\citeauthoryear{{Gugliucci} et~al.,}{{Gugliucci}
  et~al.}{2014a}]{gugliucci14a}
{Gugliucci} N.,  et~al., 2014a, in {Manning} J.~G.,  {Hemenway} M.~K.,
  {Jensen} J.~B.,   {Gibbs} M.~G.,  eds,  Astronomical Society of the Pacific
  Conference Series Vol. 483, Ensuring Stem Literacy: A National Conference on
  STEM Education and Public Outreach. p.~237

\bibitem[\protect\citeauthoryear{{Gugliucci}, {Gay}  \& {Bracey}}{{Gugliucci}
  et~al.}{2014b}]{gugliucci14b}
{Gugliucci} N.,  {Gay} P.,   {Bracey} G.,  2014b, in {Manning} J.~G.,
  {Hemenway} M.~K.,  {Jensen} J.~B.,   {Gibbs} M.~G.,  eds,  Astronomical
  Society of the Pacific Conference Series Vol. 483, Ensuring Stem Literacy: A
  National Conference on STEM Education and Public Outreach. p.~437

\bibitem[\protect\citeauthoryear{{Head} \& {Lloyd}}{{Head} \&
  {Lloyd}}{1972}]{head72}
{Head} J. W.,  {Lloyd} D.~D.,  1972, {"Near-terminator photography", in Apollo
  15: Preliminary Science Report, NASA SP-289, 25-95}.
 Vol. 289

\bibitem[\protect\citeauthoryear{{Hirabayashi}, {Minton}  \&
  {Fassett}}{{Hirabayashi} et~al.}{2017}]{hirabashi17}
{Hirabayashi} M.,  {Minton} D.~A.,   {Fassett} C.~I.,  2017, \mn@doi [\icarus]
  {10.1016/j.icarus.2016.12.032}, \href
  {https://ui.adsabs.harvard.edu/abs/2017Icar..289..134H} {289, 134}

\bibitem[\protect\citeauthoryear{Hunter}{Hunter}{2007}]{Hunter:2007}
Hunter J.~D.,  2007, \mn@doi [Computing in Science \& Engineering]
  {10.1109/MCSE.2007.55}, 9, 90

\bibitem[\protect\citeauthoryear{Jones, Oliphant  \& Peterson}{Jones
  et~al.}{2001}]{jones01}
Jones E.,  Oliphant T.,   Peterson P.,  2001

\bibitem[\protect\citeauthoryear{{Joy} et~al.,}{{Joy} et~al.}{2011}]{joy11}
{Joy} K.,  et~al., 2011, \mn@doi [Astronomy and Geophysics]
  {10.1111/j.1468-4004.2011.52210.x}, \href
  {https://ui.adsabs.harvard.edu/abs/2011A&G....52b..10J} {52, 2.10}

\bibitem[\protect\citeauthoryear{{Kreslavsky}, {Head}, {Neumann}, {Rosenburg},
  {Aharonson}, {Smith}  \& {Zuber}}{{Kreslavsky} et~al.}{2013}]{kreslavsky13}
{Kreslavsky} M.~A.,  {Head} J.~W.,  {Neumann} G.~A.,  {Rosenburg} M.~A.,
  {Aharonson} O.,  {Smith} D.~E.,   {Zuber} M.~T.,  2013, \mn@doi [\icarus]
  {10.1016/j.icarus.2013.04.027}, \href
  {https://ui.adsabs.harvard.edu/abs/2013Icar..226...52K} {226, 52}

\bibitem[\protect\citeauthoryear{{Marshall}, {Lintott}  \&
  {Fletcher}}{{Marshall} et~al.}{2015}]{marshall15}
{Marshall} P.~J.,  {Lintott} C.~J.,   {Fletcher} L.~N.,  2015, \mn@doi [\araa]
  {10.1146/annurev-astro-081913-035959}, \href
  {http://adsabs.harvard.edu/abs/2015ARA%26A..53..247M} {53, 247}

\bibitem[\protect\citeauthoryear{{Moore}}{{Moore}}{1972}]{moore72}
{Moore} H. J.,  1972, {"Crater-shadowing effects at low Sun angle", in Apollo
  15: Preliminary Science Report, NASA SP-289, 25-92}.
 Vol. 289

\bibitem[\protect\citeauthoryear{Neish, Blewett, Harmon, Coman, Cahill  \&
  Ernst}{Neish et~al.}{2013}]{neish13}
Neish C.~D.,  Blewett D.~T.,  Harmon J.~K.,  Coman E.~I.,  Cahill J. T.~S.,
  Ernst C.~M.,  2013, \mn@doi [Journal of Geophysical Research: Planets]
  {10.1002/jgre.20166}, 118, 2247

\bibitem[\protect\citeauthoryear{{Neukum} \& {Ivanov}}{{Neukum} \&
  {Ivanov}}{1994}]{neukum94}
{Neukum} G.,  {Ivanov} B.~A.,  1994, in {Gehrels} T.,  {Matthews} M.~S.,
  {Schumann} A.~M.,  eds, Hazards Due to Comets and Asteroids. p.~359

\bibitem[\protect\citeauthoryear{{Neukum}, {Ivanov}  \& {Hartmann}}{{Neukum}
  et~al.}{2001}]{neukum01}
{Neukum} G.,  {Ivanov} B.~A.,   {Hartmann} W.~K.,  2001, \ssr, \href
  {http://adsabs.harvard.edu/abs/2001SSRv...96...55N} {96, 55}

\bibitem[\protect\citeauthoryear{{Ostrach}, {Robinson}, {Denevi}  \&
  {Thomas}}{{Ostrach} et~al.}{2011}]{ostrach11}
{Ostrach} L.~R.,  {Robinson} M.~S.,  {Denevi} B.~W.,   {Thomas} P.~C.,  2011,
  in Lunar and Planetary Science Conference. p.~1202

\bibitem[\protect\citeauthoryear{Raddick, Bracey, Gay, Lintott, Murray,
  Schawinski, Szalay  \& Vandenberg}{Raddick et~al.}{2010}]{Raddick_2010}
Raddick M.~J.,  Bracey G.,  Gay P.~L.,  Lintott C.~J.,  Murray P.,  Schawinski
  K.,  Szalay A.~S.,   Vandenberg J.,  2010, \mn@doi [Astronomy Education
  Review] {10.3847/aer2009036}, 9

\bibitem[\protect\citeauthoryear{{Robbins}, {Antonenko}, {Gay}, {Lehan}  \&
  {Moore}}{{Robbins} et~al.}{2012}]{robbins12}
{Robbins} S.~J.,  {Antonenko} I.,  {Gay} P.~L.,  {Lehan} C.,   {Moore} J.,
  2012, in Lunar and Planetary Science Conference. Lunar and Planetary Science
  Conference.
p.~2856

\bibitem[\protect\citeauthoryear{{Robbins} et~al.,}{{Robbins}
  et~al.}{2014}]{robbins14}
{Robbins} S.~J.,  et~al., 2014, \mn@doi [\icarus]
  {10.1016/j.icarus.2014.02.022}, \href
  {http://adsabs.harvard.edu/abs/2014Icar..234..109R} {234, 109}

\bibitem[\protect\citeauthoryear{Robbins, Riggs, Weaver, Bierhaus, Chapman,
  Kirchoff, Singer  \& Gaddis}{Robbins et~al.}{2018}]{robbins18}
Robbins S.~J.,  Riggs J.~D.,  Weaver B.~P.,  Bierhaus E.~B.,  Chapman C.~R.,
  Kirchoff M.~R.,  Singer K.~N.,   Gaddis L.~R.,  2018, \mn@doi [Meteoritics \&
  Planetary Science] {https://doi.org/10.1111/maps.12990}, 53, 891

\bibitem[\protect\citeauthoryear{{Robinson} et~al.,}{{Robinson}
  et~al.}{2010}]{robinson10}
{Robinson} M.~S.,  et~al., 2010, \mn@doi [\ssr] {10.1007/s11214-010-9634-2},
  \href {http://adsabs.harvard.edu/abs/2010SSRv..150...81R} {150, 81}

\bibitem[\protect\citeauthoryear{{Schultz}, {Greeley}  \& {Gault}}{{Schultz}
  et~al.}{1977}]{schultz77}
{Schultz} P.~H.,  {Greeley} R.,   {Gault} D.,  1977, in {Merril} R.~B.,  ed.,
  Lunar and Planetary Science Conference Proceedings Vol. 8, Lunar and
  Planetary Science Conference Proceedings. pp 3539--3564

\bibitem[\protect\citeauthoryear{{Shoemaker} \& {Morris}}{{Shoemaker} \&
  {Morris}}{1970}]{shoemaker70}
{Shoemaker} E.~M.,  {Morris} E.~C.,  1970, \mn@doi [Radio Science]
  {10.1029/RS005i002p00129}, \href
  {http://adsabs.harvard.edu/abs/1970RaSc....5..129S} {5, 129}

\bibitem[\protect\citeauthoryear{{Soderblom}}{{Soderblom}}{1972}]{soderblom72}
{Soderblom} L.,  1972, National Aeronautics and Space Administration Special,
  pp 87--91

\bibitem[\protect\citeauthoryear{{Sutherland} \& W}{{Sutherland} \&
  W}{1974}]{sutherland74}
{Sutherland} I.~E.,  W H.~G.,  1974, Communications of the ACM, 17, 32

\bibitem[\protect\citeauthoryear{{Wilcox}, {Robinson}, {Thomas}  \&
  {Hawke}}{{Wilcox} et~al.}{2005}]{wilcox05}
{Wilcox} B.~B.,  {Robinson} M.~S.,  {Thomas} P.~C.,   {Hawke} B.~R.,  2005,
  \mn@doi [Meteoritics and Planetary Science]
  {10.1111/j.1945-5100.2005.tb00974.x}, \href
  {http://adsabs.harvard.edu/abs/2005M%26PS...40..695W} {40, 695}

\bibitem[\protect\citeauthoryear{{Wilhelms}, {McCauley}  \& {Trask}}{{Wilhelms}
  et~al.}{1987}]{wilhelms87}
{Wilhelms} D.~E.,  {McCauley} J.~F.,   {Trask} N.~J.,  1987, {The geologic
  history of the moon}

\bibitem[\protect\citeauthoryear{{Xiao} \& {Werner}}{{Xiao} \&
  {Werner}}{2015}]{xiao15}
{Xiao} Z.,  {Werner} S.~C.,  2015, \mn@doi [Journal of Geophysical Research
  (Planets)] {10.1002/2015JE004860}, \href
  {http://adsabs.harvard.edu/abs/2015JGRE..120.2277X} {120, 2277}

\bibitem[\protect\citeauthoryear{{Young}}{{Young}}{1975}]{young75}
{Young} R.~A.,  1975, in Lunar and Planetary Science Conference Proceedings. pp
  2645--2662

\bibitem[\protect\citeauthoryear{{van der Walt}, {Colbert}  \&
  {Varoquaux}}{{van der Walt} et~al.}{2011}]{2011CSE....13b..22V}
{van der Walt} S.,  {Colbert} S.~C.,   {Varoquaux} G.,  2011, \mn@doi
  [Computing in Science and Engineering] {10.1109/MCSE.2011.37}, \href
  {https://ui.adsabs.harvard.edu/abs/2011CSE....13b..22V} {13, 22}

\makeatother
\end{thebibliography}

\newpage
\appendix

\begin{table*}[!htbp]
  \scriptsize
  \begin{tabular}{*{8}{l}}
    \hline
    Diameter (km) & Area (km$^{2}$) & \multicolumn{3}{c}{Cumulative} & \multicolumn{3}{c}{R Plot} \\
    \cline{3-5}
    \cline{6-8}
    & & N$_\text{cum}$ & Density & Uncertainty & n & Relative & Uncertainty \\
    \hline
    1.00 $\times$ 10$^{-2}$ & 6.68 & 1028/1135 & 153.84/169.85 & 4.8/5.04  & 619/699 & 3.76 $\times$ 10$^{-2}$/4.25 $\times$ 10$^{-2}$ & 1.51 $\times$ 10$^{-3}$/1.61 $\times$ 10$^{-3}$\\
    1.41 $\times$ 10$^{-2}$ & 6.68 & 409/436   & 61.21/65.25   & 3.03/3.12 & 257/264 & 3.12 $\times$ 10$^{-2}$/3.21 $\times$ 10$^{-2}$ & 1.95 $\times$ 10$^{-3}$/1.97 $\times$ 10$^{-3}$\\
    2.00 $\times$ 10$^{-2}$ & 6.68 & 152/172   & 22.75/25.74   & 1.84/1.96 & 107/119 & 2.60 $\times$ 10$^{-2}$/2.89 $\times$ 10$^{-2}$ & 2.51 $\times$ 10$^{-3}$/2.65 $\times$ 10$^{-3}$\\
    2.83 $\times$ 10$^{-2}$ & 6.68 & 45/53     & 6.73/7.93     & 1.0/1.09  & 28/34   & 1.36 $\times$ 10$^{-2}$/1.65 $\times$ 10$^{-2}$ & 2.57 $\times$ 10$^{-3}$/2.83 $\times$ 10$^{-3}$\\
    4.00 $\times$ 10$^{-2}$ & 6.68 & 17/19     & 2.54/2.84     & 0.62/0.65 & 13/14   & 1.26 $\times$ 10$^{-2}$/1.36 $\times$ 10$^{-2}$ & 3.51 $\times$ 10$^{-3}$/3.64 $\times$ 10$^{-3}$\\
    5.66 $\times$ 10$^{-2}$ & 6.68 & 4/5       & 0.6/0.75      & 0.3/0.33  & 3/4     & 5.83 $\times$ 10$^{-3}$/7.78 $\times$ 10$^{-3}$ & 3.37 $\times$ 10$^{-3}$/3.89 $\times$ 10$^{-3}$\\
    8.00 $\times$ 10$^{-2}$ & 6.68 & 1/1       & 0.15/0.15     & 0.15/0.15 & 1/1     & 3.89 $\times$ 10$^{-3}$/3.89 $\times$ 10$^{-3}$ & 3.89 $\times$ 10$^{-3}$/3.89 $\times$ 10$^{-3}$\\
    \hline
  \end{tabular}
  
  \caption{Crater population data for region compared between NAC images  {\tt {M109215691RE}} ($27.5 ^{\circ}$) and  {\tt {M111578606RE}} ($38.0 ^{\circ}$). Column one gives the lower edge value of each bin. Bin widths conform to 2$^{\frac{1}{2}}$ bin intervals.  Column two gives the surface area on which the data were measured. The next three columns are the cratering statistics for the cumulative size-frequency distributions. Each column has been split in two to show values derived from the overlap region for  {\tt {M109215691RE}} and  {\tt {M111578606RE}}, respectively. The third column is the cumulative number of craters. The fourth column corresponds to the density, which is in units of cumulative number of craters per square kilometer. The fifth column yields the corresponding 1$\sigma$ uncertainty for each density value. The remaining three columns are the cratering statistics for the relative size-frequency distributions. Each column has been split in two to show values derived from the overlap region for  {\tt {M109215691RE}} and  {\tt {M111578606RE}}, respectively. The sixth column is the number of craters per bin. The seventh column corresponds to the relative frequency of craters per bin, which is a unitless quantity. The eighth column yields the 1$\sigma$ uncertainty for each R value. {The R value is calculated with respect to the geometric mean of the bin, rather than the bin minimum diameter.}}
  \label{tab:one}
\end{table*}

\begin{table*}[!htbp]
  \scriptsize
  \begin{tabular}{*{8}{l}}
    \hline
    Diameter (km) & Area (km$^{2}$) & \multicolumn{3}{c}{Cumulative} & \multicolumn{3}{c}{R Plot} \\
    \cline{3-5}
    \cline{6-8}
    & & N$_\text{cum}$ & Density & Uncertainty & n & Relative & Uncertainty \\
    \hline
    1.00 $\times$ 10$^{-2}$ & 9.14 & 1563/2024 & 171.06/221.51 & 4.33/4.92 & 968/1211 & 4.30 $\times$ 10$^{-2}$/5.38 $\times$ 10$^{-2}$ & 1.38 $\times$ 10$^{-3}$/1.55 $\times$ 10$^{-3}$\\
    1.41 $\times$ 10$^{-2}$ & 9.14 & 595/813   & 65.12/88.98   & 2.67/3.12 & 388/503 & 3.45 $\times$ 10$^{-2}$/4.47 $\times$ 10$^{-2}$  & 1.75 $\times$ 10$^{-3}$/1.99 $\times$ 10$^{-3}$\\
    2.00 $\times$ 10$^{-2}$ & 9.14 & 207/310   & 22.65/33.93   & 1.57/1.93 & 134/185 & 2.38 $\times$ 10$^{-2}$/3.29 $\times$ 10$^{-2}$  & 2.06 $\times$ 10$^{-3}$/2.42 $\times$ 10$^{-3}$\\
    2.83 $\times$ 10$^{-2}$ & 9.14 & 73/125    & 7.99/13.68    & 0.94/1.22 & 52/73   & 1.85 $\times$ 10$^{-2}$/2.59 $\times$ 10$^{-2}$  & 2.56 $\times$ 10$^{-3}$/3.04 $\times$ 10$^{-3}$\\
    4.00 $\times$ 10$^{-2}$ & 9.14 & 21/52     & 2.3/5.69      & 0.5/0.79  & 13/36   & 9.24 $\times$ 10$^{-3}$/2.56 $\times$ 10$^{-2}$  & 2.56 $\times$ 10$^{-3}$/4.27 $\times$ 10$^{-3}$\\
    5.66 $\times$ 10$^{-2}$ & 9.14 & 8/16      & 0.88/1.75     & 0.31/0.44 & 5/10    & 7.11 $\times$ 10$^{-3}$/1.42 $\times$ 10$^{-2}$  & 3.18 $\times$ 10$^{-3}$/4.50 $\times$ 10$^{-3}$\\
    8.00 $\times$ 10$^{-2}$ & 9.14 & 3/6       & 0.33/0.66     & 0.19/0.27 & 3/6     & 8.53 $\times$ 10$^{-3}$/1.71 $\times$ 10$^{-2}$  & 4.93 $\times$ 10$^{-3}$/6.97 $\times$ 10$^{-3}$\\
    \hline
  \end{tabular}
  
  \caption{Crater population data for region compared between NAC images  {\tt {M109215691LE/RE}} ($27.5 ^{\circ}$) and  {\tt {M119829425LE}} ($58.0 ^{\circ}$). For description of columns, see Table~\ref{tab:one}.}
\end{table*}

\begin{table*}[!htbp]
  \scriptsize
  \begin{tabular}{*{8}{l}}
    \hline
    Diameter (km) & Area (km$^{2}$) & \multicolumn{3}{c}{Cumulative} & \multicolumn{3}{c}{R Plot} \\
    \cline{3-5}
    \cline{6-8}
    & & N$_\text{cum}$ & Density & Uncertainty & n & Relative & Uncertainty \\
    \hline
    1.00 $\times$ 10$^{-2}$ & 14.16 & 2541/3969 & 179.39/280.21 & 3.56/4.45 & 1544/2368 & 4.43 $\times$ 10$^{-2}$/6.79 $\times$ 10$^{-2}$ & 1.13 $\times$ 10$^{-3}$/1.39 $\times$ 10$^{-3}$\\
    1.41 $\times$ 10$^{-2}$ & 14.16 & 997/1601  & 70.39/113.03  & 2.23/2.82 & 633/897   & 3.63 $\times$ 10$^{-2}$/5.14 $\times$ 10$^{-2}$ & 1.44 $\times$ 10$^{-3}$/1.72 $\times$ 10$^{-3}$\\
    2.00 $\times$ 10$^{-2}$ & 14.16 & 364/704   & 25.7/49.7     & 1.35/1.87 & 233/362   & 2.67 $\times$ 10$^{-2}$/4.15 $\times$ 10$^{-2}$ & 1.75 $\times$ 10$^{-3}$/2.18 $\times$ 10$^{-3}$\\
    2.83 $\times$ 10$^{-2}$ & 14.16 & 131/342   & 9.25/24.14    & 0.81/1.31 & 83/181    & 1.90 $\times$ 10$^{-2}$/4.15 $\times$ 10$^{-2}$ & 2.09 $\times$ 10$^{-3}$/3.09 $\times$ 10$^{-3}$\\
    4.00 $\times$ 10$^{-2}$ & 14.16 & 48/161    & 3.39/11.37    & 0.49/0.9  & 28/92     & 1.28 $\times$ 10$^{-2}$/4.22 $\times$ 10$^{-2}$ & 2.43 $\times$ 10$^{-3}$/4.40 $\times$ 10$^{-3}$\\
    5.66 $\times$ 10$^{-2}$ & 14.16 & 20/69     & 1.41/4.87     & 0.32/0.59 & 14/43     & 1.28 $\times$ 10$^{-2}$/3.94 $\times$ 10$^{-2}$ & 3.43 $\times$ 10$^{-3}$/6.02 $\times$ 10$^{-3}$\\
    8.00 $\times$ 10$^{-2}$ & 14.16 & 6/26      & 0.42/1.84     & 0.17/0.36 & 6/26      & 1.10 $\times$ 10$^{-2}$/4.77 $\times$ 10$^{-2}$ & 4.49 $\times$ 10$^{-3}$/9.35 $\times$ 10$^{-3}$\\
    \hline
  \end{tabular}
  
  \caption{Crater population data for region compared between NAC images  {\tt {M109215691LE/RE}} ($27.5 ^{\circ}$) and  {\tt {M146959973LE/RE}} ($77.5 ^{\circ}$). For description of columns, see Table~\ref{tab:one}.}
\end{table*}

\begin{table*}[!htbp]
  \scriptsize
  \begin{tabular}{*{8}{l}}
    \hline
    Diameter (km) & Area (km$^{2}$) & \multicolumn{3}{c}{Cumulative} & \multicolumn{3}{c}{R Plot} \\
    \cline{3-5}
    \cline{6-8}
    & & N$_\text{cum}$ & Density & Uncertainty & n & Relative & Uncertainty \\
    \hline
    1.00 $\times$ 10$^{-2}$ & 8.4 & 1450/2361 & 172.7/281.2 & 4.54/5.79 & 901/1403 & 4.36 $\times$ 10$^{-2}$/6.78 $\times$ 10$^{-2}$ & 1.45 $\times$ 10$^{-3}$/1.81 $\times$ 10$^{-3}$\\
    1.41 $\times$ 10$^{-2}$ & 8.4 & 549/958   & 65.39/114.1 & 2.79/3.69 & 353/552 & 3.41 $\times$ 10$^{-2}$/5.34 $\times$ 10$^{-2}$  & 1.82 $\times$ 10$^{-3}$/2.27 $\times$ 10$^{-3}$\\
    2.00 $\times$ 10$^{-2}$ & 8.4 & 196/406   & 23.34/48.36 & 1.67/2.4  & 125/221 & 2.42 $\times$ 10$^{-2}$/4.27 $\times$ 10$^{-2}$  & 2.16 $\times$ 10$^{-3}$/2.88 $\times$ 10$^{-3}$\\
    2.83 $\times$ 10$^{-2}$ & 8.4 & 71/185    & 8.46/22.03  & 1.0/1.62  & 51/109 & 1.97 $\times$ 10$^{-2}$/4.22 $\times$ 10$^{-2}$   & 2.76 $\times$ 10$^{-3}$/4.04 $\times$ 10$^{-3}$\\
    4.00 $\times$ 10$^{-2}$ & 8.4 & 20/76     & 2.38/9.05   & 0.53/1.04 & 12/38 & 9.28 $\times$ 10$^{-3}$/2.94 $\times$ 10$^{-2}$    & 2.68 $\times$ 10$^{-3}$/4.77 $\times$ 10$^{-3}$\\
    5.66 $\times$ 10$^{-2}$ & 8.4 & 8/38      & 0.95/4.53   & 0.34/0.73 & 4/26 & 6.19 $\times$ 10$^{-3}$/4.02 $\times$ 10$^{-2}$     & 3.09 $\times$ 10$^{-3}$/7.89 $\times$ 10$^{-3}$\\
    8.00 $\times$ 10$^{-2}$ & 8.4 & 4/12      & 0.48/1.43   & 0.24/0.41 & 4/12 & 1.24 $\times$ 10$^{-2}$/3.71 $\times$ 10$^{-2}$     & 6.19 $\times$ 10$^{-3}$/1.07 $\times$ 10$^{-2}$\\
    \hline
  \end{tabular}
  
  \caption{Crater population data for region compared between NAC images  {\tt {M109215691LE/RE}} ($27.5 ^{\circ}$) and  {\tt {M117467833RE}} ($83.0 ^{\circ}$). For description of columns, see Table~\ref{tab:one}.}
\end{table*}

\begin{table*}[!htbp]
  \scriptsize
  \begin{tabular}{*{8}{l}}
    \hline
    Diameter (km) & Area (km$^{2}$) & \multicolumn{3}{c}{Cumulative} & \multicolumn{3}{c}{R Plot} \\
    \cline{3-5}
    \cline{6-8}
    & & N$_\text{cum}$ & Density & Uncertainty & n & Relative & Uncertainty \\
    \hline
    1.00 $\times$ 10$^{-2}$ & 5.12 & 905/1121 & 176.76/218.95 & 5.88/6.54 & 562/673 & 4.46 $\times$ 10$^{-2}$/5.34 $\times$ 10$^{-2}$ & 1.88 $\times$ 10$^{-3}$/2.06 $\times$ 10$^{-3}$\\
    1.41 $\times$ 10$^{-2}$ & 5.12 & 343/448  & 66.99/87.5    & 3.62/4.13 & 206/271 & 3.27 $\times$ 10$^{-2}$/4.30 $\times$ 10$^{-2}$ & 2.28 $\times$ 10$^{-3}$/2.61 $\times$ 10$^{-3}$\\
    2.00 $\times$ 10$^{-2}$ & 5.12 & 137/177  & 26.76/34.57   & 2.29/2.6  & 91/113  & 2.89 $\times$ 10$^{-2}$/3.58 $\times$ 10$^{-2}$ & 3.03 $\times$ 10$^{-3}$/3.37 $\times$ 10$^{-3}$\\
    2.83 $\times$ 10$^{-2}$ & 5.12 & 46/64    & 8.98/12.5     & 1.32/1.56 & 32/37   & 2.03 $\times$ 10$^{-2}$/2.35 $\times$ 10$^{-2}$ & 3.59 $\times$ 10$^{-3}$/3.86 $\times$ 10$^{-3}$\\
    4.00 $\times$ 10$^{-2}$ & 5.12 & 14/27    & 2.73/5.27     & 0.73/1.01 & 9/20    & 1.14 $\times$ 10$^{-2}$/2.54 $\times$ 10$^{-2}$ & 3.81 $\times$ 10$^{-3}$/5.67 $\times$ 10$^{-3}$\\
    5.66 $\times$ 10$^{-2}$ & 5.12 & 5/7      & 0.98/1.37     & 0.44/0.52 & 4/6     & 1.02 $\times$ 10$^{-2}$/1.52 $\times$ 10$^{-2}$ & 5.08 $\times$ 10$^{-3}$/6.22 $\times$ 10$^{-3}$\\
    8.00 $\times$ 10$^{-2}$ & 5.12 & 1/1      & 0.2/0.2       & 0.2/0.2   & 1/1     & 5.08 $\times$ 10$^{-3}$/5.08 $\times$ 10$^{-3}$ & 5.08 $\times$ 10$^{-3}$/5.08 $\times$ 10$^{-3}$\\
    \hline
  \end{tabular}
  
  \caption{Crater population data for region compared between NAC images  {\tt {M111578606RE}} ($38.0 ^{\circ}$) and  {\tt {M119829425LE}} ($58.0 ^{\circ}$). For description of columns, see Table~\ref{tab:one}.}
\end{table*}

\begin{table*}[!htbp]
  \scriptsize
  \begin{tabular}{*{8}{l}}
    \hline
    Diameter (km) & Area (km$^{2}$) & \multicolumn{3}{c}{Cumulative} & \multicolumn{3}{c}{R Plot} \\
    \cline{3-5}
    \cline{6-8}
    & & N$_\text{cum}$ & Density & Uncertainty & n & Relative & Uncertainty \\
    \hline
    1.00 $\times$ 10$^{-2}$ & 6.48 & 1117/1481 & 172.35/228.52 & 5.16/5.94 & 689/862 & 4.32 $\times$ 10$^{-2}$/5.40 $\times$ 10$^{-2}$ & 1.64 $\times$ 10$^{-3}$/1.84 $\times$ 10$^{-3}$\\
    1.41 $\times$ 10$^{-2}$ & 6.48 & 428/619   & 66.04/95.51 & 3.19/3.84   & 260/359 & 3.26 $\times$ 10$^{-2}$/4.50 $\times$ 10$^{-2}$ & 2.02 $\times$ 10$^{-3}$/2.37 $\times$ 10$^{-3}$\\
    2.00 $\times$ 10$^{-2}$ & 6.48 & 168/260   & 25.92/40.12 & 2.0/2.49    & 117/154 & 2.93 $\times$ 10$^{-2}$/3.86 $\times$ 10$^{-2}$ & 2.71 $\times$ 10$^{-3}$/3.11 $\times$ 10$^{-3}$\\
    2.83 $\times$ 10$^{-2}$ & 6.48 & 51/106    & 7.87/16.36 & 1.1/1.59     & 33/50 & 1.65 $\times$ 10$^{-2}$/2.51 $\times$ 10$^{-2}$   & 2.88 $\times$ 10$^{-3}$/3.54 $\times$ 10$^{-3}$\\
    4.00 $\times$ 10$^{-2}$ & 6.48 & 18/56     & 2.78/8.64 & 0.65/1.15     & 14/33 & 1.40 $\times$ 10$^{-2}$/3.31 $\times$ 10$^{-2}$   & 3.75 $\times$ 10$^{-3}$/5.76 $\times$ 10$^{-3}$\\
    5.66 $\times$ 10$^{-2}$ & 6.48 & 4/23      & 0.62/3.55 & 0.31/0.74     & 3/16 & 6.01 $\times$ 10$^{-3}$/3.21 $\times$ 10$^{-2}$    & 3.47 $\times$ 10$^{-3}$/8.02 $\times$ 10$^{-3}$\\
    8.00 $\times$ 10$^{-2}$ & 6.48 & 1/7       & 0.15/1.08 & 0.15/0.41     & 1/7 & 4.01 $\times$ 10$^{-3}$/2.81 $\times$ 10$^{-2}$     & 4.01 $\times$ 10$^{-3}$/1.06 $\times$ 10$^{-2}$\\
    \hline
  \end{tabular}
  
  \caption{Crater population data for region compared between NAC images  {\tt {M111578606RE}} ($38.0 ^{\circ}$) and  {\tt {M146959973LE/RE}} ($77.5 ^{\circ}$). For description of columns, see Table~\ref{tab:one}.}
\end{table*}

\begin{table*}[!htbp]
  \scriptsize
  \begin{tabular}{*{8}{l}}
    \hline
    Diameter (km) & Area (km$^{2}$) & \multicolumn{3}{c}{Cumulative} & \multicolumn{3}{c}{R Plot} \\
    \cline{3-5}
    \cline{6-8}
    & & N$_\text{cum}$ & Density & Uncertainty & n & Relative & Uncertainty \\
    \hline
    1.00 $\times$ 10$^{-2}$ & 4.59 & 823/1285 & 179.24/279.86 & 6.25/7.81 & 500/771 & 4.42 $\times$ 10$^{-2}$/6.82 $\times$ 10$^{-2}$ & 1.98 $\times$ 10$^{-3}$/2.46 $\times$ 10$^{-3}$\\
    1.41 $\times$ 10$^{-2}$ & 4.59 & 323/514  & 70.35/111.94  & 3.91/4.94 & 192/291 & 3.40 $\times$ 10$^{-2}$/5.15 $\times$ 10$^{-2}$ & 2.45 $\times$ 10$^{-3}$/3.02 $\times$ 10$^{-3}$\\
    2.00 $\times$ 10$^{-2}$ & 4.59 & 131/223  & 28.53/48.57   & 2.49/3.25 & 86/128  & 3.04 $\times$ 10$^{-2}$/4.53 $\times$ 10$^{-2}$ & 3.28 $\times$ 10$^{-3}$/4.00 $\times$ 10$^{-3}$\\
    2.83 $\times$ 10$^{-2}$ & 4.59 & 45/95    & 9.8/20.69     & 1.46/2.12 & 31/50   & 2.19 $\times$ 10$^{-2}$/3.54 $\times$ 10$^{-2}$ & 3.94 $\times$ 10$^{-3}$/5.00 $\times$ 10$^{-3}$\\
    4.00 $\times$ 10$^{-2}$ & 4.59 & 14/45    & 3.05/9.8      & 0.81/1.46 & 9/24    & 1.27 $\times$ 10$^{-2}$/3.40 $\times$ 10$^{-2}$ & 4.24 $\times$ 10$^{-3}$/6.93 $\times$ 10$^{-3}$\\
    5.66 $\times$ 10$^{-2}$ & 4.59 & 5/21     & 1.09/4.57     & 0.49/1.0  & 4/16    & 1.13 $\times$ 10$^{-2}$/4.53 $\times$ 10$^{-2}$ & 5.66 $\times$ 10$^{-3}$/1.13 $\times$ 10$^{-2}$\\
    8.00 $\times$ 10$^{-2}$ & 4.59 & 1/5      & 0.22/1.09     & 0.22/0.49 & 1/5     & 5.66 $\times$ 10$^{-3}$/2.83 $\times$ 10$^{-2}$ & 5.66 $\times$ 10$^{-3}$/1.27 $\times$ 10$^{-2}$\\
    \hline
  \end{tabular}
  
  \caption{Crater population data for region compared between NAC images  {\tt {M111578606RE}} ($38.0 ^{\circ}$) and  {\tt {M117467833RE}} ($83.0 ^{\circ}$). For description of columns, see Table~\ref{tab:one}.}
\end{table*}

\begin{table*}[!htbp]
  \scriptsize
  \begin{tabular}{*{8}{l}}
    \hline
    Diameter (km) & Area (km$^{2}$) & \multicolumn{3}{c}{Cumulative} & \multicolumn{3}{c}{R Plot} \\
    \cline{3-5}
    \cline{6-8}
    & & N$_\text{cum}$ & Density & Uncertainty & n & Relative & Uncertainty \\
    \hline
    1.00 $\times$ 10$^{-2}$ & 9.04 & 2006/2520 & 221.79/278.62 & 4.95/5.55 & 1201/1523 & 5.39 $\times$ 10$^{-2}$/6.84 $\times$ 10$^{-2}$ & 1.56 $\times$ 10$^{-3}$/1.75 $\times$ 10$^{-3}$\\
    1.41 $\times$ 10$^{-2}$ & 9.04 & 805/997   & 89.0/110.23   & 3.14/3.49 & 499/553   & 4.48 $\times$ 10$^{-2}$/4.96 $\times$ 10$^{-2}$ & 2.01 $\times$ 10$^{-3}$/2.11 $\times$ 10$^{-3}$\\
    2.00 $\times$ 10$^{-2}$ & 9.04 & 306/444   & 33.83/49.09   & 1.93/2.33 & 182/236   & 3.27 $\times$ 10$^{-2}$/4.24 $\times$ 10$^{-2}$ & 2.42 $\times$ 10$^{-3}$/2.76 $\times$ 10$^{-3}$\\
    2.83 $\times$ 10$^{-2}$ & 9.04 & 124/208   & 13.71/23.0    & 1.23/1.59 & 73/113    & 2.62 $\times$ 10$^{-2}$/4.06 $\times$ 10$^{-2}$ & 3.07 $\times$ 10$^{-3}$/3.82 $\times$ 10$^{-3}$\\
    4.00 $\times$ 10$^{-2}$ & 9.04 & 51/95     & 5.64/10.5     & 0.79/1.08 & 35/54     & 2.51 $\times$ 10$^{-2}$/3.88 $\times$ 10$^{-2}$ & 4.25 $\times$ 10$^{-3}$/5.28 $\times$ 10$^{-3}$\\
    5.66 $\times$ 10$^{-2}$ & 9.04 & 16/41     & 1.77/4.53     & 0.44/0.71 & 10/29     & 1.44 $\times$ 10$^{-2}$/4.17 $\times$ 10$^{-2}$ & 4.54 $\times$ 10$^{-3}$/7.74 $\times$ 10$^{-3}$\\
    8.00 $\times$ 10$^{-2}$ & 9.04 & 6/12      & 0.66/1.33     & 0.27/0.38 & 6/12      & 1.72 $\times$ 10$^{-2}$/3.45 $\times$ 10$^{-2}$ & 7.04 $\times$ 10$^{-3}$/9.95 $\times$ 10$^{-3}$\\
    \hline
  \end{tabular}
  
  \caption{Crater population data for region compared between NAC images  {\tt {M119829425LE}} ($58.0 ^{\circ}$) and  {\tt {M146959973LE/RE}} ($77.5 ^{\circ}$). For description of columns, see Table~\ref{tab:one}.}
\end{table*}

\begin{table*}[!htbp]
  \scriptsize
  \begin{tabular}{*{8}{l}}
    \hline
    Diameter (km) & Area (km$^{2}$) & \multicolumn{3}{c}{Cumulative} & \multicolumn{3}{c}{R Plot} \\
    \cline{3-5}
    \cline{6-8}
    & & N$_\text{cum}$ & Density & Uncertainty & n & Relative & Uncertainty \\
    \hline
    1.00 $\times$ 10$^{-2}$ & 8.34 & 2348/1869 & 281.61/224.16 & 5.81/5.19 & 1393/1119 & 6.78 $\times$ 10$^{-2}$/5.45 $\times$ 10$^{-2}$ & 1.82 $\times$ 10$^{-3}$/1.63 $\times$ 10$^{-3}$\\
    1.41 $\times$ 10$^{-2}$ & 8.34 & 955/750   & 114.54/89.95  & 3.71/3.28 & 554/460   & 5.40 $\times$ 10$^{-2}$/4.48 $\times$ 10$^{-2}$ & 2.29 $\times$ 10$^{-3}$/2.09 $\times$ 10$^{-3}$\\
    2.00 $\times$ 10$^{-2}$ & 8.34 & 401/290   & 48.09/34.78   & 2.4/2.04  & 219/170   & 4.27 $\times$ 10$^{-2}$/3.31 $\times$ 10$^{-2}$ & 2.88 $\times$ 10$^{-3}$/2.54 $\times$ 10$^{-3}$\\
    2.83 $\times$ 10$^{-2}$ & 8.34 & 182/120   & 21.83/14.39   & 1.62/1.31 & 106/72    & 4.13 $\times$ 10$^{-2}$/2.80 $\times$ 10$^{-2}$ & 4.01 $\times$ 10$^{-3}$/3.31 $\times$ 10$^{-3}$\\
    4.00 $\times$ 10$^{-2}$ & 8.34 & 76/48     & 9.12/5.76     & 1.05/0.83 & 38/32     & 2.96 $\times$ 10$^{-2}$/2.49 $\times$ 10$^{-2}$ & 4.80 $\times$ 10$^{-3}$/4.41 $\times$ 10$^{-3}$\\
    5.66 $\times$ 10$^{-2}$ & 8.34 & 38/16     & 4.56/1.92     & 0.74/0.48 & 26/11     & 4.05 $\times$ 10$^{-2}$/1.71 $\times$ 10$^{-2}$ & 7.95 $\times$ 10$^{-3}$/5.17 $\times$ 10$^{-3}$\\
    8.00 $\times$ 10$^{-2}$ & 8.34 & 12/5      & 1.44/0.6      & 0.42/0.27 & 12/5      & 3.74 $\times$ 10$^{-2}$/1.56 $\times$ 10$^{-2}$ & 1.08 $\times$ 10$^{-2}$/6.97 $\times$ 10$^{-3}$\\
    \hline
  \end{tabular}
  
  \caption{Crater population data for region compared between NAC images  {\tt {M117467833RE}} ($83.0 ^{\circ}$) and  {\tt {M119829425LE}} ($58.0 ^{\circ}$). For description of columns, see Table~\ref{tab:one}.}
\end{table*}

\begin{table*}[!htbp]
  \scriptsize
  \begin{tabular}{*{8}{l}}
    \hline
    Diameter (km) & Area (km$^{2}$) & \multicolumn{3}{c}{Cumulative} & \multicolumn{3}{c}{R Plot} \\
    \cline{3-5}
    \cline{6-8}
    & & N$_\text{cum}$ & Density & Uncertainty & n & Relative & Uncertainty \\
    \hline
    1.00 $\times$ 10$^{-2}$ & 8.35 & 2343/2347 & 280.63/281.11 & 5.8/5.8   & 1392/1418 & 6.77 $\times$ 10$^{-2}$/6.90 $\times$ 10$^{-2}$ & 1.81 $\times$ 10$^{-3}$/1.83 $\times$ 10$^{-3}$\\
    1.41 $\times$ 10$^{-2}$ & 8.35 & 951/929   & 113.9/111.27  & 3.69/3.65 & 550/507   & 5.35 $\times$ 10$^{-2}$/4.93 $\times$ 10$^{-2}$ & 2.28 $\times$ 10$^{-3}$/2.19 $\times$ 10$^{-3}$\\
    2.00 $\times$ 10$^{-2}$ & 8.35 & 401/422   & 48.03/50.54   & 2.4/2.46  & 220/225   & 4.28 $\times$ 10$^{-2}$/4.38 $\times$ 10$^{-2}$ & 2.89 $\times$ 10$^{-3}$/2.92 $\times$ 10$^{-3}$\\
    2.83 $\times$ 10$^{-2}$ & 8.35 & 181/197   & 21.68/23.6    & 1.61/1.68 & 106/108   & 4.12 $\times$ 10$^{-2}$/4.20 $\times$ 10$^{-2}$ & 4.01 $\times$ 10$^{-3}$/4.04 $\times$ 10$^{-3}$\\
    4.00 $\times$ 10$^{-2}$ & 8.35 & 75/89     & 8.98/10.66    & 1.04/1.13 & 37/50     & 2.88 $\times$ 10$^{-2}$/3.89 $\times$ 10$^{-2}$ & 4.73 $\times$ 10$^{-3}$/5.50 $\times$ 10$^{-3}$\\
    5.66 $\times$ 10$^{-2}$ & 8.35 & 38/39     & 4.55/4.67     & 0.74/0.75 & 26/27     & 4.05 $\times$ 10$^{-2}$/4.20 $\times$ 10$^{-2}$ & 7.94 $\times$ 10$^{-3}$/8.09 $\times$ 10$^{-3}$\\
    8.00 $\times$ 10$^{-2}$ & 8.35 & 12/12     & 1.44/1.44     & 0.41/0.41 & 12/12     & 3.73 $\times$ 10$^{-2}$/3.73 $\times$ 10$^{-2}$ & 1.08 $\times$ 10$^{-2}$/1.08 $\times$ 10$^{-2}$\\
    \hline
  \end{tabular}
  
  \caption{Crater population data for region compared between NAC images  {\tt {M117467833RE}} ($83.0 ^{\circ}$) and  {\tt {M146959973LE/RE}} ($77.5 ^{\circ}$). For description of columns, see Table~\ref{tab:one}.}
\end{table*}
\end{document}